%% file: irs_vf.tex
\documentclass[journal]{IEEEtran}
\usepackage{stmaryrd}
\usepackage{amssymb,amsmath,amsfonts,wasysym,latexsym}
\usepackage[acronyms,shortcuts]{glossaries}
\usepackage[none]{hyphenat}
\usepackage{bm}
\usepackage{tikz}
\usetikzlibrary{patterns,arrows,decorations.pathreplacing}
\usepackage{psfrag}
\usepackage{amssymb}

\usepackage{epstopdf}
\usepackage{amsmath}
\usepackage{scalefnt}
\usepackage[none]{hyphenat}
\usepackage{color}
\usepackage{algorithm}
\usepackage{algpseudocode}
\usepackage{cite}
\usepackage{caption}
\usepackage{subcaption}

\usepackage{caption}
\usepackage{graphicx}

\definecolor{forestgreen}{rgb}{0, 0.5,0.5} 
\definecolor{darkgreen}{rgb}{0,0.392157,0} 
\newcommand{\nmode}[2]{\left[\ma{\mathcal{#1}}\right]_{\left(#2\right)}}
\newcommand{\sumx}[2]{\sum\limits_{#1}^{#2}}
\newcommand{\bb}[1]{\mathbb{#1}}
\newcommand{\ten}[1]{\boldsymbol{\mathcal #1}}
\newcommand{\ma}[1]{\boldsymbol{#1}}
\definecolor{green}{rgb}{0.1,0.75,0.2}

\input{acronym-irs}

\begin{document}
\title{ 
Reducing the Control Overhead of Intelligent Reconfigurable Surfaces Via a Tensor-Based Low-Rank Factorization Approach
 } 
\author{Bruno Sokal,~\IEEEmembership{Student Member,~IEEE,} Paulo R. B. Gomes, André L. F. de Almeida,~\IEEEmembership{Senior Member,~IEEE,} Behrooz Makki,~\IEEEmembership{Senior Member,~IEEE,} and Gabor Fodor,~\IEEEmembership{Senior Member,~IEEE}
        
     \thanks{Bruno Sokal, Paulo R. B. Gomes, and Andr\'{e} L. F. de Almeida are with the Wireless Telecom Research Group (GTEL), Department of Teleinformatics Engineering, Federal University of Cear\'{a}, Fortaleza-CE, Brazil. E-mails: \{brunosokal,andre\}@gtel.ufc.br.}
\thanks{Behrooz Makki is with  Ericsson Research, Göteborg, Sweden . E-mail: behrooz.makki@ericsson.com}
\thanks{Gabor Fodor is with  Ericsson Research and KTH Royal Institute of Technology, Stockholm, Sweden . E-mail: gabor.fodor@ericsson.com}
\thanks{This work was supported by the Ericsson Research, Sweden, and Ericsson Innovation Center, Brazil, under UFC.48 Technical Cooperation Contract Ericsson/UFC. This study was financed in part by the Coordenação de Aperfeiçoamento de
Pessoal de Nível Superior - Brasil (CAPES)-Finance Code
001, and CAPES/PRINT Proc. 88887.311965/2018-00. Andr\'{e}~L.~F.~de~Almeida acknowledges CNPq for its financial support under the grant 312491/2020-4. G. Fodor was partially supported by the Digital Futures project PERCy.

\textcolor{black}{Part of this work has been submitted for possible presentation in IEEE GLOBECOM 2022 \cite{bs_globecom_arkv}. } } 
}

\maketitle

\begin{abstract}

\textcolor{black}{Passive  intelligent reconfigurable surfaces (IRS) are becoming an attractive component of cellular networks due to their ability of shaping the propagation environment and thereby improving the coverage. While passive IRS nodes incorporate a great number of phase-shifting elements and a controller entity, the phase-shifts are typically determined by the cellular base station (BS) due to its computational capability. Since the fine granularity control of the large number of phase-shifters may become prohibitive in practice, it is important to reduce the control overhead between the BS and the IRS controller. To this end, in this paper we propose a low-rank approximation of the near-optimal phase-shifts, which would incur prohibitively high communication overhead on the BS-IRS controller links.  \textcolor{black}{The key idea is to represent the potentially large IRS phase-shift vector using a low-rank tensor model. This is achieved by factorizing a \emph{tensorized} version of the IRS phase-shift vector, where each component is modeled as the Kronecker product of a predefined number of factors of smaller sizes, which can be  obtained \emph{via} tensor decomposition algorithms. We show that the proposed low-rank models  drastically reduce  the required feedback requirements associated with the BS-IRS control links. } Our simulation results indicate that the proposed method is especially attractive in scenarios with a strong line of sight component, in which case nearly the same spectral efficiency is reached as in the cases with near-optimal phase-shifts, but with a drastically reduced communication overhead.} 
\end{abstract}


\section{Introduction}\label{Sec:Introduction}
\Ac{IRS} is a  \textcolor{black}{candidate} technology for beyond fifth generation and sixth generation networks  due to its ability to \textcolor{black}{\textit{control}} the electromagnetic properties of the radio-frequency  waves by performing an intelligent phase-shift to the desired direction \cite{zhang_tutorial,Gong_Survey,jian2022reconfigurable,di2020smart,basar2019,Basar2020,ozdogan2019intelligent,rajatheva2020scoring}. Usually,  \ac{IRS} is defined as a planar ($2$-D) surface with a large number of independent reflective elements, in which they can be fully passive or with some elements active \cite{Taha2019,khoshafa2021active,alexandropoulos2020hardware}. \ac{IRS} is  connected to a smart controller  that sets the desired phase-shift for each reflective element, by applying  bias voltages at the elements e.g., PIN diodes. The main advantage of  fully passive \acp{IRS} is its full-duplex nature, i.e., no noise amplification is observed since no signal processing is possible. However, \textcolor{black}{the fully passive nature of the IRSs} \textcolor{black}{makes} the \ac{CSI} acquisition process \textcolor{black}{difficult}, since no pilots are processed, thus only the cascade channel can be estimated.  Nevertheless, in the case of employing a few active elements in the \ac{IRS}, this issue is suppressed and channel can be estimated using, for example, compressed sensing tools \cite{Taha2019}. Another advantage of an \ac{IRS} with fully passive elements is that the power consumption is concentrated  at the controller. \textcolor{black}{This makes the \ac{IRS}} a more attractive technology in terms of  energy efficiency compared to \textcolor{black}{alternative technologies}, \textcolor{black}{e.g.,} amplify-and-forward and decode-and-forward relays \cite{Huang2019,emil2019_relay,guo2022dynamic}.

Several works have addressed the \ac{CSI} acquisition problem in IRS-assisted networks, e.g., \cite{gil2021,gil2022,Chen2019,Hu2019,li2020joint,ardah2021trice,an2021low,2021Wei}. The work of \cite{gil2021} proposes a tensor-based method where the authors show the benefits of exploiting the multidimensional structure of the received signal by separating the cascade channel.  The work of \cite{Chen2019} proposes a compressed sensing approach in a multi-user uplink \ac{MIMO} scenario. In \cite{Hu2019}, a two-timescale channel estimation framework is proposed to overcome the pilot overhead in a multi-user \ac{IRS}-aided system. \textcolor{black}{Also, \cite{li2020joint} addresses} the channel estimation problem in millimiter-wave  \ac{MIMO} systems. 
\textcolor{black}{The work of \cite{ardah2021trice} proposes a channel estimation framework for \ac{mmWave} \ac{IRS}-assisted \ac{MIMO} systems based on compressed sensing techniques. The authors of \cite{an2021low} propose  a low-complexity framework for channel estimation and passive beamforming in \ac{MIMO} IRS-assisted systems.}

\textcolor{black}{Although many works  focus on channel estimation \cite{gil2021,gil2022,Chen2019,Hu2019,li2020joint,ardah2021trice,an2021low,2021Wei}, achievable rate maximization \cite{2021Wei,yang2020intelligent,2021Perovic}, \textcolor{black}{\ac{EE}
maximization \cite{gao2021outage,du2021reconfigurable,you2020energy,chen2022energy,siddiqi2022energy}},  and interference mitigation problems \cite{khaleel2021novel,ntougias2021interference,elmossallamy2021ris},} few works have addressed the \textcolor{black}{problem of reducing the channel training or the feedback-overhead of IRS phase-shifts to the IRS controller.} 
The work of \cite{yang2020intelligent} proposes a protocol design to maximize the transmission rate in IRS-assisted \ac{MIMO}-OFDM systems. Also, \cite{Zappone_Overhead_Aware} proposes a framework for overhead-aware feedback  and resource allocation in \ac{IRS}-assisted \ac{MIMO} systems. \textcolor{black}{The main idea of \cite{Zappone_Overhead_Aware} is to optimize the network resource such as the bandwidth and the total power used for transmission and feedback. However, the number of phase-shifts to be conveyed to the IRS controller can still be large and results in high feedback signaling overhead, especially for large IRS panels.} 

In this work, we propose a overhead-aware model for designing the IRS phase-shifts. \textcolor{black}{Our idea is to represent the IRS phase-shift vector with a low-rank model. This is achieved by factorizing a \emph{tensorized} version of the IRS phase-shift vector, where each component is modelled as the Kronecker product of a predefined number of factors. } These factors are estimated using  tensor decompositions such as the \ac{PARAFAC} \cite{PARAFAC} and Tucker \cite{tucker}. After the estimation process, the phases   of the factors are quantized and fed back to the IRS controller, which can reconstruct the IRS phase-shift vector based on the chosen low-rank tensor model. The main contributions of this work are the following:

\begin{enumerate}
{\color{black}
    \item Our proposed IRS phase-shift factorization allows to save network resources by reducing the total IRS phase-shift feedback overhead. This allows a more frequent IRS phase-shift feedback, for a fixed feedback load, which can significantly improve the end-to-end latency, crucial in a fast varying channels, high mobility scenarios and/or the cases with moderate/large sizes of the IRS. Also, thanks to the significant reduction on the feedback overhead, the IRS-assisted network can decide to multiplex phase-shifts associated with a higher number of users in the same feedback channel. 
    \item  The proposed IRS phase-shift factorization provides a flexible feedback design by controlling the parameters of the low-rank factorization model, such as the number of components, the number and the size of the factors, as well as their respective resolution. This is an important  feature of our proposed feedback-aware model, since for limited feedback control links, the low-rank model and its factorization parameters can be efficiently  adjusted to the available capacity of the feedback link, providing more degrees of freedom to system design.
    \item  Our tensor-based factorization approach relies on the optimum IRS phase-shift vector, which means that it can be implemented in \textcolor{black}{every} IRS-assisted network and in \textcolor{black}{multiple} communication links, i.e.,  downlink or uplink, in single-input single-output, multiple-input single-output, \textcolor{black}{as well in} \ac{MIMO} systems.
    
%
}
\end{enumerate}

Different from the works of \cite{yang2020intelligent} and \cite{Zappone_Overhead_Aware}, we aim \textcolor{black}{to reduce}  the IRS phase-shifts feedback \textcolor{black}{overhead} by conveying to the IRS controller only the factors of our proposed low-rank model. Our approach is analytical and provides a systematic way of controlling the feedback overhead by adjusting the parameters of the low-rank IRS model, namely, its rank and the corresponding number of factors of each rank-one component.  Our simulations show that the proposed low-rank model for the IRS phase-shifts can achieve the same \ac{SE} as the state-of-the-art in \ac{LOS} scenarios, while the feedback payload (number of bits to be fed back) is dramatically reduced. For example, taking an IRS with $N=1024$ elements, the feedback duration can be $50$ times smaller than the state-of-the-art, depending on the low-rank model parameters. Also, when taking into account the \textcolor{black}{total} system \ac{SE} and \ac{EE}, i.e., both the IRS phase-shift feedback duration, and the channel estimation duration, our proposed model outperforms the state-of-the-art. 

The rest of the paper is organized as follows. {\color{black} Section \ref{Sec:Tensor_background} provides an introduction of the tensor notation and decompositions that are exploited in this paper. The system model is described in Section \ref{Sec:System_Model}. Section \ref{Sec:Proposed_Method} details our proposed feedback overhead-aware method and provides the details of the PARAFAC-IRS and Tucker-IRS models for IRS phase-shift vector factorization. Section \ref{eq:quan_rec_par}  describes the phase-shift and weighting factors quantization procedure and the reconstruction of the IRS phase-shift vector at the IRS controller. The effects of the factorization parameters and the quantization process are also discussed in this section. Simulation results are provided in Section \ref{Sec:Simulation_Results} and the final conclusions and perspectives are discussed in Section \ref{Sec:conclusions}.
}

\subsection{Notation and Properties} \label{Sec:notation}
Scalars are represented as non-bold lower-case letters $a$, column vectors as lower-case boldface letters $\ma{a}$, matrices as upper-case boldface letters $\ma{A}$, and tensors as calligraphic upper-case letters $\ten{A}$. The superscripts $\{\cdot\}^{\text{T}}$, $\{\cdot\}^{\text{*}}$, $\{\cdot\}^{\text{H}}$ and $\{\cdot\}^{+ }$ stand for transpose, conjugate, conjugate transpose and pseudo-inverse operations, respectively. The operator $\Arrowvert\cdot\Arrowvert_{\text{F}}$ denotes the Frobenius norm of a matrix or tensor, and $\bb{E}\{\cdot\}$ is the expectation operator. The operator $\text{diag}\left(\ma{a}\right)$ converts $\ma{a}$ into a diagonal matrix, while $\text{diag}(\ma{A})$ returns a vector whose elements are the main the diagonal of $\ma{A}$.  Moreover, $\text{vec}\left(\ma{A}\right)$ converts $\ma{A} \in \mathbb{C}^{I_{1} \times R}$ to a column vector $\ma{a} \in \mathbb{C}^{I_{1}R \times 1}$ by stacking its columns on top of each other, while the unvec($\cdot$) operator is the inverse  of the vec operation. The symbol $\circ$ denotes the outer product operator. Also, \textcolor{black}{$\ma{a}_{r} \in \mathbb{C}^{I \times 1}$ represents the $r$-th column of $\ma{A} \in \bb{C}^{I \times R}$}. Let us define  two matrices $\ma{A} = \left[\ma{a}_{1}, \ldots, \ma{a}_{R}\right] \in \mathbb{C}^{I_{1} \times R}$ and $\ma{B} = \left[\ma{b}_{1}, \ldots, \ma{b}_{R}\right] \in \mathbb{C}^{I_{2} \times R}$. The  Kronecker product between them is defined by
\begin{align*}
\ma{A} \otimes \ma{B} = \left[ \begin{array}{ccc}
a_{1,1}\ma{B}  &  \ldots  &  a_{1,R}\ma{B} \\
\vdots         & \ddots  &  \vdots \\
a_{I,1}\ma{B}         &   &  a_{I,R}\ma{B}  
    \end{array} \right] \in \bb{C}^{I_2I_1 \times R R}.
\end{align*}

The Khatri-Rao product, also known as the column wise Kronecker product, between two matrices, symbolized by $\diamond$, is defined as 
\begin{align*}
\ma{A} \diamond \ma{B} = \left[\ma{a}_{1} \otimes \ma{b}_{1}, \ldots, \ma{a}_{R} \otimes \ma{b}_{R}\right] \in \mathbb{C}^{I_{2}I_{1} \times R}.
\end{align*}

We make use of the following properties
\begin{align}
\label{p1}\text{vec}\left(\ma{A}\ma{B}\ma{C}\right) &= \left(\ma{C}^{\text{T}} \otimes \ma{A}\right)\text{vec}\left(\ma{B}\right), \\
\label{p2} \text{vec}\left(\ma{A}\text{diag}\left( \ma{b}\right)\ma{C}\right) &= \left(\ma{C}^{\text{T}} \diamond \ma{A}\right)\ma{b},\\
\label{p3}\ma{a}^{\text{T}} \diamond \ma{B} &= \ma{B}\text{diag}(\ma{a}),\\
\label{p4}\ma{a} \otimes \ma{b} &= \text{vec}\left(\ma{b} \circ \ma{a}\right),
\end{align}
where the involved vectors and matrices  have compatible dimensions in each case.
\section{Tensor Pre-Requisites}\label{Sec:Tensor_background}

In this section, tensor preliminaries are provided by focusing on the main notation, operations and properties that will be useful in the rest of the paper. 

Consider a set of matrices $\{\ma{X}_{i_3}\} \in \bb{C}^{I_1 \times I_2}$, for $i_3 = 1,\ldots, I_3$. Concatenating all $I_3$ matrices, we form the third-order tensor $\ten{X} = [\ma{X}_1 \sqcup_3 \ma{X}_2 \sqcup_3 \ldots \sqcup_3 \ma{X}_{I_3} ] \in \bb{C}^{I_1 \times I_2 \times I_3}$, where $\sqcup_3 $ indicates a concatenation in the third dimension. We can interpret $\ma{X}_{i_3}$ as the $i_3$-th frontal slice of $\ten{X}$, defined as $\ten{X}_{..i_3} = \ma{X}_{i_3}$ where the ``$..$" indicates that the dimensions $I_1$ and $I_2$ are fixed. The tensor $\ten{X}$ can be \textcolor{black}{\textit{matricized}} by letting one dimension vary along the rows and the remaining two dimensions along the columns. From $\ten{X}$, we can form three different matrices, referred to as the $n$-mode unfoldings (for $n=\{1,2,3 \}$ in this case), which are respectively given by
\begin{align}
\label{eq:nmode_1}\nmode{X}{1} &= [\ten{X}_{..1},\ldots,\ten{X}_{..I_3}] \in \bb{C}^{I_1 \times I_2I_3}, \\  
\label{eq:nmode_2}\nmode{X}{2} &= [\ten{X}_{..1}^{\text{T}},\ldots,\ten{X}_{..I_3}^{\text{T}}] \in \bb{C}^{I_2 \times I_1I_3}\\ 
\label{eq:nmode_3}\nmode{X}{3} &= [\text{vec}(\ten{X}_{..1}),\ldots,\text{vec}(\ten{X}_{..I_3})]^{\text{T}} \in \bb{C}^{I_3 \times I_1I_2}.
\end{align} 

\subsection{Tensorization}\label{Sec:tensor_background_tensorization}
{\color{black}
The tensorization operation consists  of mapping the elements of a vector into a high-order tensor. Let us define the vector $\ma{y} \in \bb{C}^{N \times 1}$, in which $N = \prod\limits_{p=1}^P {N_p}$, where $N_p$ is the size of the $p$-th partition of this vector. By applying the tensorization operator, defined as $\ten{T} \{ \cdot \}$, we can form a $P$-order tensor $\ten{Y} = \ten{T}\{\ma{y}\}  \in \bb{C}^{N_1 \times N_2 \times \ldots \times N_P}$. } The mapping of elements from $\ma{y}$ to $\ten{Y}$ is defined as
\begin{align}
\label{eq:tensozire}
\ten{Y}_{n_1,n_2,\ldots, n_P} = \ma{y}_{n_1 + (n_2 -1)N_1  + \ldots + (n_P - 1)N_{P-1}\cdots N_2N_1},
\end{align}
where $n_p = \{1,\ldots , N_p \}$, for $p = \{1, \ldots ,P\}$. {\color{black} This operator plays a key role on the proposed method, and will be exploited to recast the IRS phase-shift vector as a tensor, from which the low-rank factorization schemes are proposed.}

\subsection{PARAFAC Decomposition}\label{Sec:PARAFAC_decom}

\textcolor{black}{It is known that every} matrix of rank $R$ can be expressed as the summation of its rank-one components obtained by, e.g., \ac{SVD}. In the case of tensors, a tensor of rank $R$ is given by the summation of its rank-one tensor factors. This decomposition is called \ac{PARAFAC} \cite{PARAFAC}. For a $P$ order tensor $\ten{Y} \in \bb{C}^{I_1 \times I_2 \times  \ldots \times I_P}$, its \ac{PARAFAC} decomposition is given as
\begin{align}
\label{eq:parafac_y_sum_R}\ten{Y} = \sumx{r=1}{R} \ma{a}^{(1)}_{r} \circ \ma{a}^{(2)}_{r} \circ  \ldots \circ \ma{a}^{(P)}_{r} \in \bb{C}^{I_1 \times I_2 \times \ldots \times I_P},
\end{align}
where $\ma{a}^{(p)}_{r} \in \bb{C}^{I_p \times 1}$ is $r$-th  column of the $p$-th factor matrix $\ma{A}^{(p)} \in \bb{C}^{I_p \times R}$, $p = \{1,\ldots, P\}$. The $p$-th mode unfolding of $\ten{Y}$, defined as $\nmode{Y}{p} \in \bb{C}^{I_p \times I_1\cdots I_{p-1}I_{p+1} \cdots I_P }$, is expressed as 
\begin{equation}
 \label{eq:nmode_parafac}   \nmode{Y}{p} = \ma{A}^{(p)}\left(\ma{A}^{(P)} \diamond \ldots \diamond \ma{A}^{(p+1)} \diamond \ma{A}^{(p-1)} \diamond \ldots \diamond \ma{A}^{(1)} \right)^{\text{T}}.
\end{equation}

\begin{figure}[!t]
	\centering\includegraphics[scale=0.085]{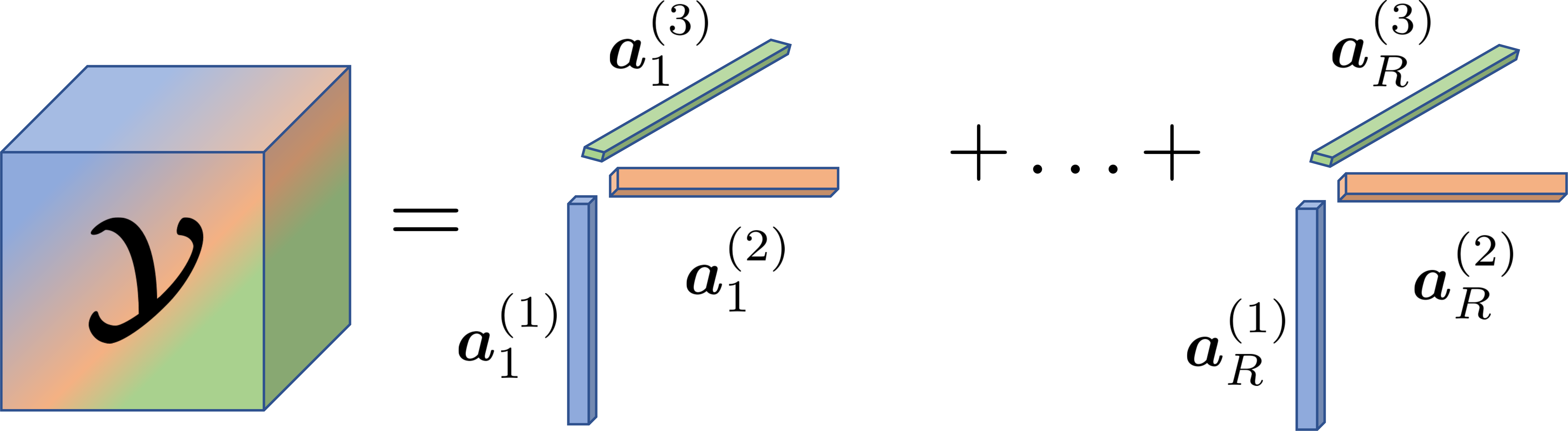}
	\caption{Illustration of a third-order PARAFAC tensor as a sum of $R$ rank-one tensors.}
	\label{fig:parafac_rank_r}
\end{figure}

Fig. \ref{fig:parafac_rank_r} illustrates a PARAFAC tensor, for $P=3$, as the summation of rank-one tensors. In this case, it can be shown that the three-mode unfoldings can be factorized as \cite{kolda}
\begin{align}
\label{eq:1_mode_parafac_y} \nmode{Y}{1} &= \ma{A}^{(1)}\left(\ma{A}^{(3)} \diamond \ma{A}^{(2)} \right)^{\text{T}} \in \bb{C}^{I_1 \times I_2I_3}, \\
\label{eq:2_mode_parafac_y}\nmode{Y}{2} &= \ma{A}^{(2)}\left(\ma{A}^{(3)} \diamond \ma{A}^{(1)} \right)^{\text{T}} \in \bb{C}^{I_2 \times I_1I_3}, \\
\label{eq:3_mode_parafac_y}\nmode{Y}{3} &= \ma{A}^{(3)}\left(\ma{A}^{(2)} \diamond \ma{A}^{(1)} \right)^{\text{T}} \in \bb{C}^{I_3 \times I_1I_2} .
\end{align}


\subsection{Tucker Decomposition}\label{Sec:Tucker_decom}

The Tucker decomposition \textcolor{black}{expresses} a tensor \textcolor{black}{as a} set of factor matrices and a core tensor.  \textcolor{black}{A} $P$-th order tensor $\ten{Q} \in \bb{C}^{I_1 \times \ldots \times I_P}$ that admits a Tucker decomposition, can be written as
\begin{align}
\ten{Q} = \ten{G} \times_1 \ma{B}^{(1)} \times_2 \ldots \times_P \ma{B}^{(P)} \in \bb{C}^{I_1 \times \ldots \times I_P},
\end{align} 
where  $\ma{B}^{(p)} \in \bb{C}^{I_p \times R_p}$ is the $p$-th factor matrix, for $p = \{1,\ldots ,P\}$, and $\ten{G} \in \bb{C}^{R_1 \times \ldots \times R_P}$ is the core tensor. 
\begin{figure}[!t]
	\centering\includegraphics[scale=0.075]{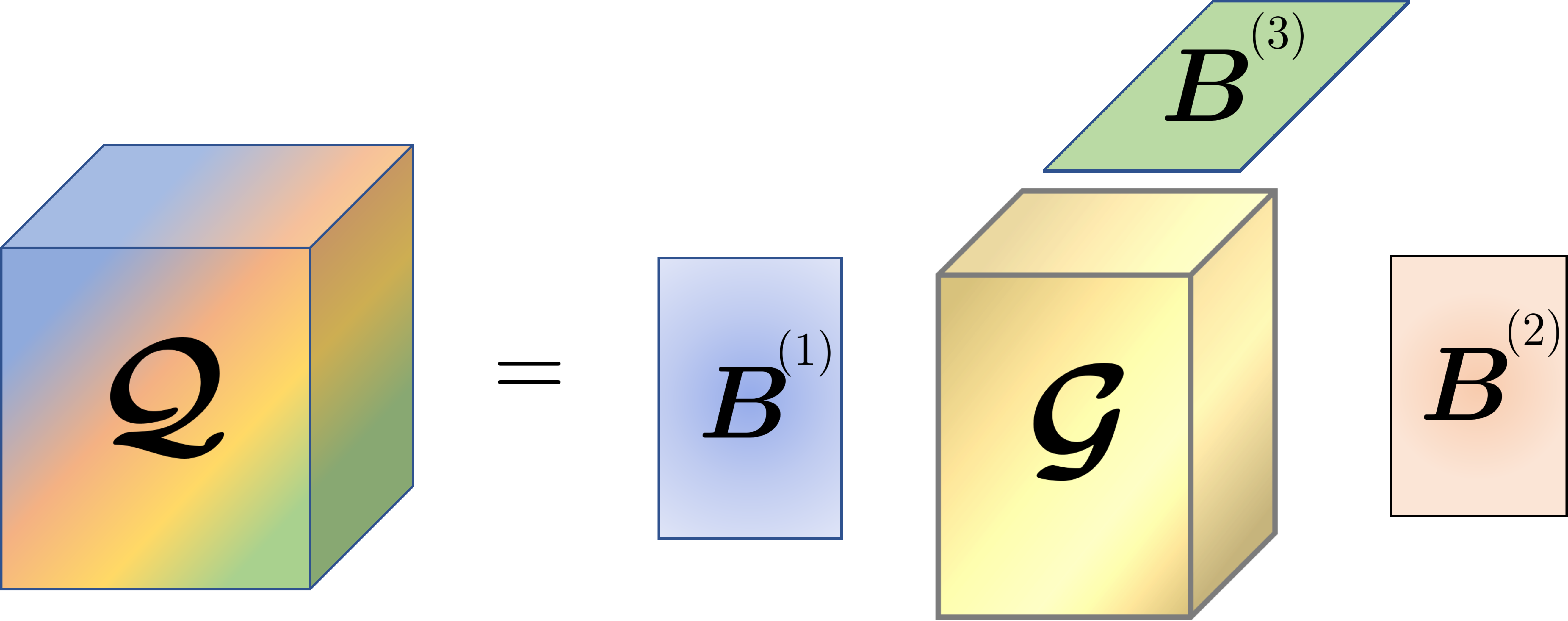}
	\caption{Illustration of a third-order Tucker tensor and its factor matrices and core tensor.}
	\label{fig:tucker_r}
\end{figure}
  \textcolor{black}{The tensor $\ten{Q}$ can also be represented as the outer product of its factors, given as}

 \begin{align*}
\ten{Q}= \sumx{r_1=1}{R_1} \ldots \sumx{r_P=1}{R_P} \ten{G}_{r_1,\ldots ,r_P} \left( \ma{b}^{(1)}_{r_1} \circ \ldots \circ \ma{b}^{(P)}_{r_P} \right),
\end{align*}
where $\ma{b}^{(p)} \in \bb{C}^{I_p \times 1}$ is the $r_p$-th column of the $p$-th factor matrix $\ma{B}^{(p)} \in \bb{C}^{I_p \times R_p}$ for $p = \{1,\ldots, P\}$ and $ r_p = \{1,\ldots, R_p\}$.  \textcolor{black}{The} $p$-th mode unfolding matrix of $\ten{Q}$, defined as $\nmode{Q}{p} \in \bb{C}^{N_p \times N_1 \cdots N_{p-1}N_{p+1} \cdots N_P}$, is given by
\begin{equation}
    \label{eq:tucker_mode_n}\nmode{Q}{p} \hspace{-0.1cm}=  \hspace{-0.1cm}\ma{B}^{(p)} \hspace{-0.07cm}\nmode{G}{p}  \hspace{-0.06cm}\left( \hspace{-0.07cm} \ma{B}^{(P)}\hspace{-0.07cm} \otimes\hspace{-0.05cm} \ldots \hspace{-0.05cm}\otimes \hspace{-0.05cm}\ma{B}^{(p+1)} \hspace{-0.05cm}\otimes \hspace{-0.05cm} \ma{B}^{(p-1)} \hspace{-0.05cm}\otimes \hspace{-0.05cm}\ldots\hspace{-0.05cm} \otimes\hspace{-0.05cm} \ma{B}^{(1)}\hspace{-0.08cm} \right)^{\text{T}}.
\end{equation}

For $P=3$, Fig. \ref{fig:tucker_r}, illustrates the decomposition. Its three mode unfoldings are given by
\begin{align}
\label{eq:1_mode_t} \nmode{Q}{1} &= \ma{B}^{(1)}\nmode{G}{1}\left(\ma{B}^{(3)} \otimes \ma{B}^{(2)} \right)^{\text{T}} \in \bb{C}^{I_1 \times I_2I_3}, \\
\label{eq:2_mode_t}\nmode{Q}{2} &= \ma{B}^{(2)}\nmode{G}{2}\left(\ma{B}^{(3)} \otimes \ma{B}^{(1)} \right)^{\text{T}} \in \bb{C}^{I_2 \times I_1I_3}, \\
\label{eq:3_mode_t}\nmode{Q}{3} &= \ma{B}^{(3)}\nmode{G}{3}\left(\ma{B}^{(2)} \otimes \ma{B}^{(1)} \right)^{\text{T}} \in \bb{C}^{I_3 \times I_1I_2}.
\end{align}

\section{System Model}\label{Sec:System_Model}
\begin{figure}[!t]
	\centering\includegraphics[scale=0.08]{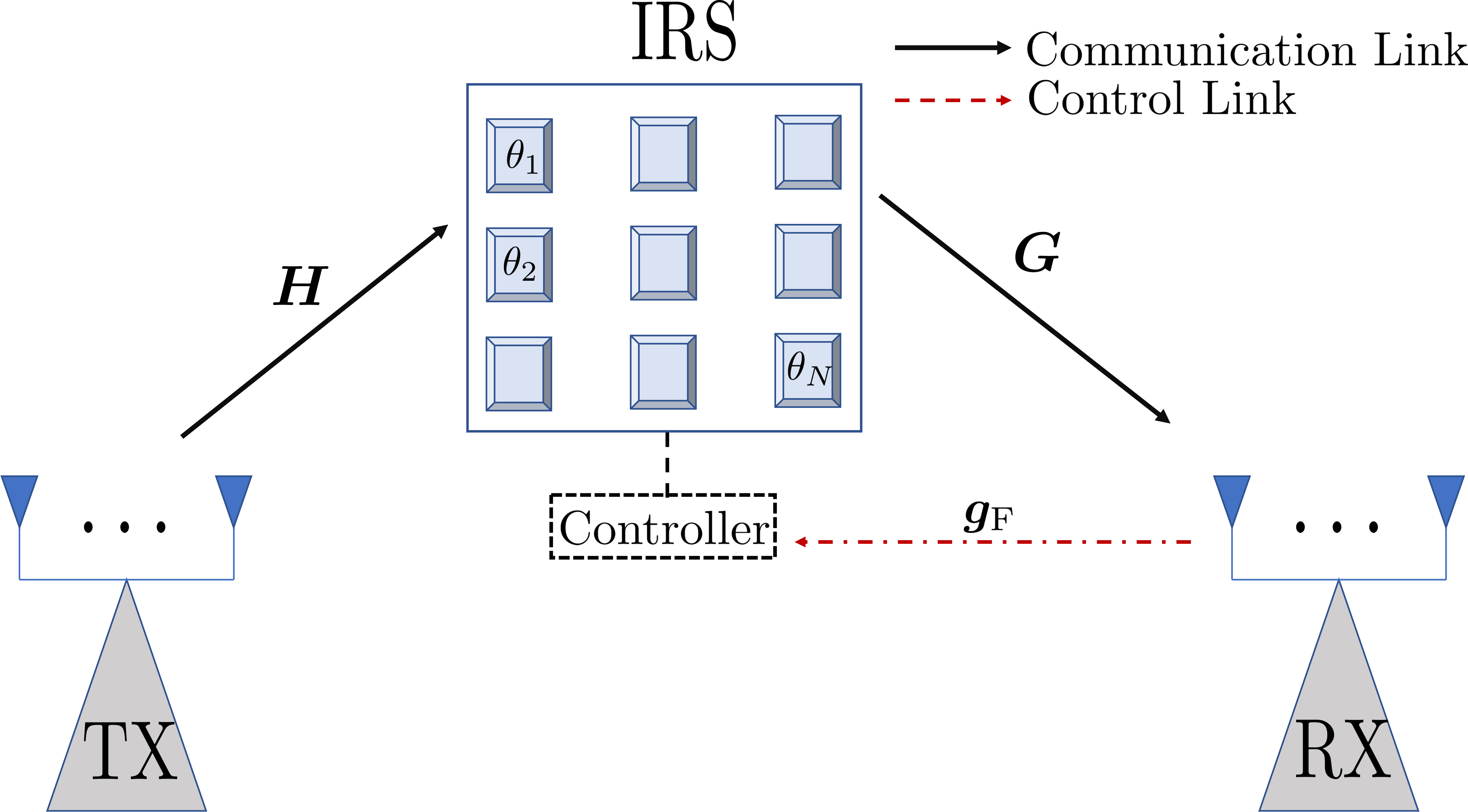}
	\caption{System model illustration.}
	\label{fig:system_model}
\end{figure}

We consider the system  illustrated in Fig. \ref{fig:system_model}, where the \ac{TX} is equipped with a \ac{ULA} with $M_T$ antenna elements, the \ac{RX} is equipped with \ac{ULA} with $M_R$ antenna elements and the \ac{IRS} has $N$ reflective elements. To simplify \textcolor{black}{the discussion}, let us consider a single stream transmission, and \textcolor{black}{assume} that there is no direct link between the \ac{TX} and \ac{RX}, \textcolor{black}{ e.g., \ac{BS}}. First, the \ac{TX} sends a pilot signal to the \ac{RX} with the aid of the \ac{IRS}. Since the \ac{IRS} has no signal processing capabilities, the channel estimation and the IRS phase-shifts optimization are performed at the \ac{RX}. The received signal after processing the pilots is given by
\begin{align}
  \label{eq:sig}  y = \ma{w}^{\text{H}}\ma{G}\ma{S}\ma{H}\ma{q} + \ma{w}^{\text{H}}\ma{b},
\end{align}
where $\ma{b} \in \bb{C}^{M_R \times 1}$ is the additive noise at the receiver with  $\mathbb{E}[\ma{b}\ma{b}^{\text{H}}] = \sigma^2_b \ma{I}_{M_r}$, $\ma{w} \in \bb{C}^{M_R \times 1}$ and $\ma{q} \in \bb{C}^{M_T \times 1}$ are the receiver and transmitter combiner and precoder, respectively. $\ma{H} \in \bb{C}^{N \times M_T}$ and $\ma{G} \in \bb{C}^{M_R \times N}$ are the TX-IRS and IRS-RX involved channels, and $\ma{S} = \text{diag}(\ma{s}) \in \bb{C}^{N \times N}$ with $\ma{s} = [e^{j\theta_1}, \ldots, e^{j\theta_N}]\in \bb{C}^{N \times 1}$ being the IRS phase-shift vector, and $\theta_n$ is the phase-shift applied to the $n$-th \ac{IRS} element. 

After the channel estimation step, the precoder and the combiner (active beamformers) vectors $\ma{w}$ and $\ma{q}$, and the IRS phase-shift vector $\ma{s}$ (passive beamformer) are optimized. Later, the \ac{RX} needs to feedback to the IRS controller the designed phase-shifts so that the \textcolor{black}{IRS controller tunes the phase-shift for each IRS element.} {\color{black} Considering the fact that this feedback occurs in a limited capacity control channel and that the IRS may contain several hundreds to thousands of reflecting elements, the feedback of each phase-shift with a certain resolution \textcolor{black}{imposes a signaling overhead.}  In this regard, the work \cite{Zappone_Overhead_Aware} models the feedback duration as}
\begin{equation}
    \label{eq:zappone_tf} T_{\text{F}} =  \frac{Nb_\text{F}}{B_{\text{F}} \text{log}\left( 1+ \frac{p_{\text{F}}|{g}_{\text{F}}|^2}{B_\text{F} N_0} \right)},
\end{equation}
where \textcolor{black}{$N$ is the total number of IRS phase-shifts to be fed back,} $B_{\text{F}}$, $p_{\text{F}}$ are the feedback bandwith and power, respectively, $g_{\text{F}}$ is the scalar control channel used, $b_{\text{F}}$ is the resolution of each phase-shift, and $N_0$ is the noise power density. The authors of \cite{Zappone_Overhead_Aware} focus on the problem of rate and \ac{EE} maximization, where the rate is given by
\begin{align}
 \label{eq:rate_zap}  \hspace{-1.5ex} \text{SE}  &= \left( 1 - \frac{T_E + T_F}{T}\right) B \text{log}\left( 1 +\frac{ p_{\text{TX}}|\ma{w}^{\text{H}}\ma{G}\ma{S}\ma{H}\ma{q}|^2 }{BN_0}\right),
\end{align}
with $T_E$ and $T$ being the duration of the channel estimation phase and the total time interval, \textcolor{black}{and $B$ the transmission bandwidth}. The \ac{EE} is given by $\text{EE} = \text{Rate}/P_{\text{tot}}$, and the total power consumption $P_{\text{tot}}$ can be expressed as 
\begin{align}
\label{eq:power_zap}\text{P}_{\text{tot}} =  P_{\text{E}} + \frac{T - T_\text{E} - T_\text{F}}{T}\mu p + \frac{\mu_\text{F} p_\text{F}T_\text{F}}{T} + P_{\text{c}},
\end{align}
where $P_\text{E}$ is the power used for the channel estimation phase, $1/\mu$ is the efficiency of the transmitter power amplifier, $p_\text{F}$ is the power used during $T_\text{F}$ seconds, and $\mu_{\text{F}}$ is the efficiency of the transmit amplifier used for feedback.  The work \cite{Zappone_Overhead_Aware} maximizes  (\ref{eq:rate_zap}) and (\ref{eq:power_zap}) by optimizing the values of the $p$, $p_{\text{F}}$, $B$, $B_\text{F}$. 


\textcolor{black}{Based on the model provided by \cite{Zappone_Overhead_Aware} in (\ref{eq:zappone_tf}), we propose to reduce the feedback overhead by factorizing the IRS phase-shift vector into smaller factors, as explained in the following section. }  



\section{Proposed Feedback-Aware Method}\label{Sec:Proposed_Method}

{\color{black}
In this section, we describe the proposed tensor low-rank approximation based feedback-aware methods that focus on reducing the feedback duration $T_\text{F}$, given in (\ref{eq:zappone_tf}). First, we assume that the \ac{RX} has an estimate of the involved channels $\ma{H}$ and $\ma{G}$.} The $N$ phase-shifts of the IRS \textcolor{black}{can be} determined based on \textcolor{black}{different} state-of-the-art algorithms (see, e.g., \cite{Zappone_Overhead_Aware,basar2019}), and are represented in a vector format as $\ma{s} = [e^{j\theta_1}, e^{j\theta_2}, \ldots, e^{j\theta_N}]  \in \bb{C}^{N \times 1}$. Our {\color{black} initial idea } consists of factorizing $\ma{s}$ as the Kronecker product of $P$ factors, i.e., 
\begin{equation}
    \label{eq:fac_s} \ma{s} =  \ma{s}^{(P)} \otimes \ldots \otimes \ma{s}^{(1)} \in \bb{C}^{N_P \cdots N_1 \times 1},
\end{equation}
where $\ma{s}^{(p)} \in \bb{C}^{N_p \times 1}$ and $N = \prod\limits_{p=1}^P N_p$. 

\begin{figure}[!t]
	\centering\includegraphics[scale=0.088]{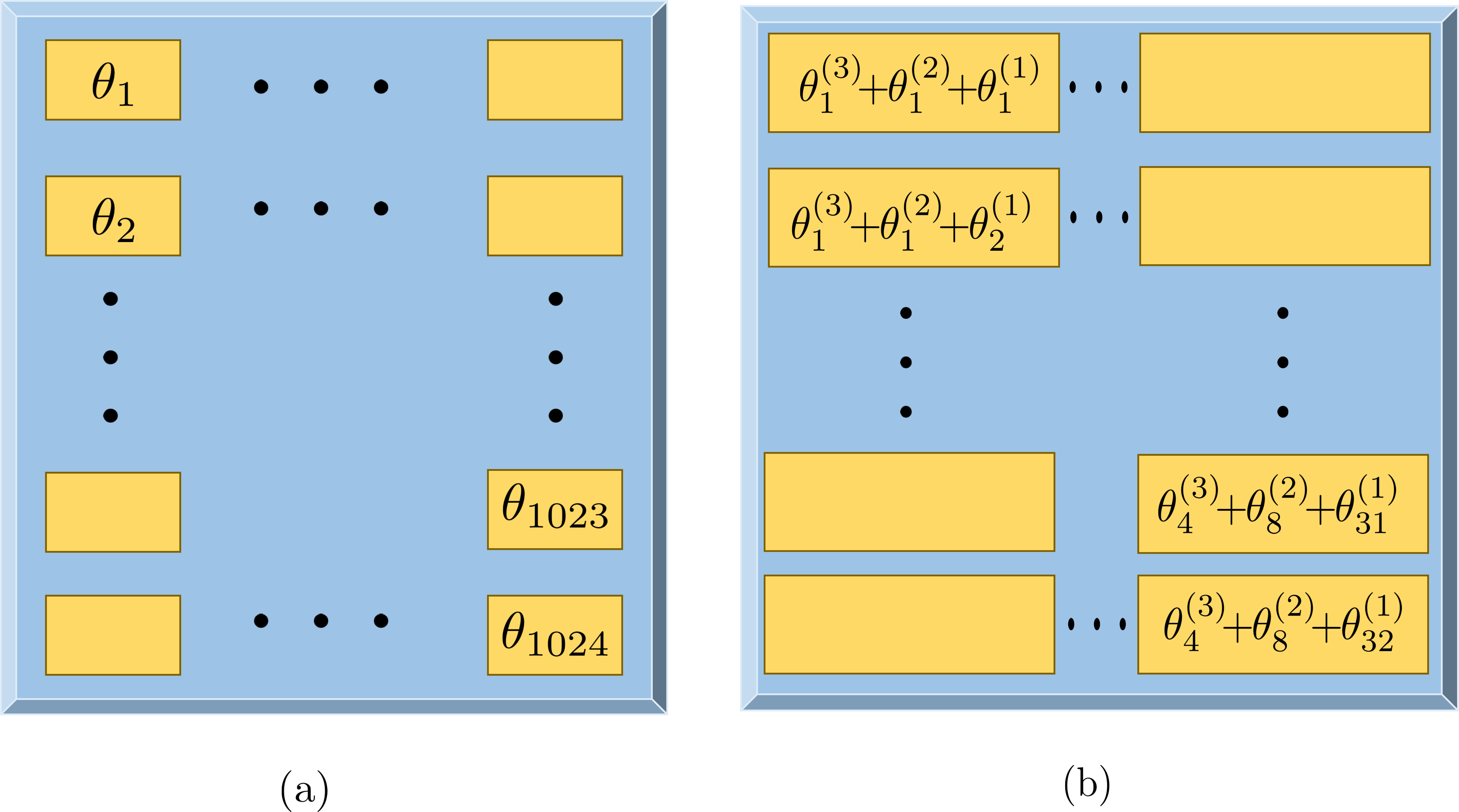}
	\caption{(a) IRS with $N=1024$ elements without factorization, (b) IRS with $N=1024$ elements factorized into $P=3$ factors. }
	\label{fig:IRS_factorized}
\end{figure}

\textbf{Example:} {\color{black} To get a first insight into } the impact of this factorization on the IRS phase-shift feedback overhead, let us consider a simple scenario with $N=1024$ phase-shifts, and we apply our factorization \textcolor{black}{method} by choosing $P=3$ factors. Consider, as one example, the following factors $\ma{s}^{(1)} = [e^{j\theta^{(1)}_1}, \ldots, e^{j\theta^{(1)}_{32}} ] \in \bb{C}^{32 \times 1}$, $\ma{s}^{(2)} = [e^{j\theta^{(2)}_1}, \ldots, e^{j\theta^{(2)}_8}] \in \bb{C}^{8 \times 1}$ and $\ma{s}^{(3)} = [e^{j\theta^{(3)}_1}, \ldots, e^{j\theta^{(3)}_4}] \in \bb{C}^{4 \times 1}$, i.e., $N_1 = 32$, $N_2 =8$ and $N_3=4$. Note that, $N_1$, $N_2$, $N_3$ can have \textcolor{black}{every} size as long $N_1\times N_2 \times N_3 = N = 1024$. In this scenario, instead of conveying to the IRS  controller $1024$ phase-shifts, we only need to convey the phase-shifts of the factors, i.e., $32 + 8 + 4 = 44$, reducing drastically  the total amount of phase-shift overhead. Physically, the Kronecker product in (\ref{eq:fac_s}) represents a summation of the factors phase-shifts. \textcolor{black}{It is clear that, in a general model for a large $N$, and based on the choice of $P$, we  have that $ \sumx{p=1}{P} N_P << N =\prod\limits_{p=1}^{P}N_P$.} The discussed example is illustrated in Fig. \ref{fig:IRS_factorized}. $\blacksquare$

\begin{figure}[!t]
	\centering\includegraphics[scale=0.088]{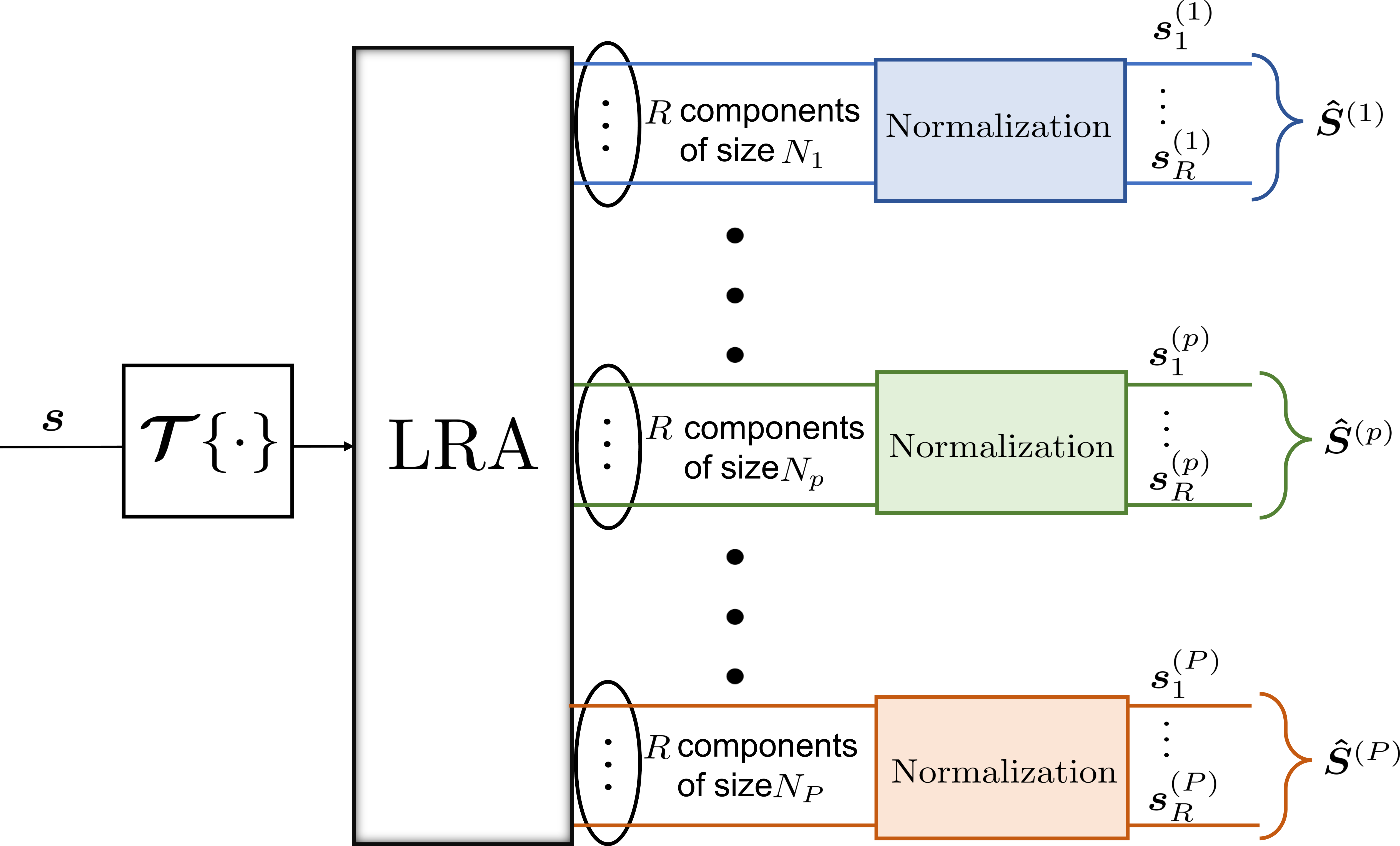}
	\caption{Proposed method for the IRS phase-shift vector factorization based on the PARAFAC-IRS model.}
	\label{fig:LRA}
\end{figure}

In a general view, the proposed factorization consists of three steps, \textcolor{black}{illustrated in Fig. \ref{fig:LRA} (for the PARAFAC-IRS model):}

\begin{enumerate}
\item \textbf{Rearrangement  of elements}: In this step, the optimum phase-shift vector $\ma{s} \in \bb{C}^{N \times 1}$ is rearranged into a $P$-th order tensor $\ten{S} \in \bb{C}^{N_1 \times N_2 \times \ldots \times N_P}$, with $N = \prod\limits_{p=1}^{P} N_p $.  This is accomplished by mapping the elements of the IRS phase-shift vector $\ma{s}$ into the tensor $\ma{S}$, using the tensorization operator, \textcolor{black}{given in   (\ref{eq:tensozire})}.
\item \textbf{\Ac{LRA}}: In this step, the \ac{RX} selects an \ac{LRA} model for $\ten{S}$ based on its unfoldings matrices.  For example, the RX can approximate the tensor $\ten{S}$ as a PARAFAC or a Tucker model, and makes use of classical tensor algorithms, such as the  \ac{ALS} \cite{kolda} and \ac{HOSVD} \cite{Lathauwer2000bestr1r2},   to estimate the factors matrices (and the core tensor, in the case of the Tucker model).   
\item \textbf{Normalization}: The factors outputs of the \ac{LRA} are normalized   due to the unitary modulus constraint of the phase-shift vector. In other words, the RX  convey to the IRS controller only the angles of the computed factors.
\end{enumerate}

In the next section, we generalize the concept of (\ref{eq:fac_s}) by factorizing the IRS phase-shift vector $\ma{s} \in \bb{C}^{N \times 1}$  based on the PARAFAC and Tucker  \ac{LRA} models.

\subsection{PARAFAC-IRS Low-Rank Approximation}\label{Sec:PARAFAC_IRS}

In the tensorization step,  a rearranging of elements from the  IRS phase-shift vector $\ma{s}$ to a tensor is performed, i.e.,  $\ten{S} = \mathcal{T}\{ \ma{s}\}$. Then, the RX will approximate the optimum phase-shift tensor $\ten{S}$ using a PARAFAC model, i.e.,
\begin{equation}
\label{eq:tenS_parafac}    \ten{S} \approx \sumx{r=1}{R} \ma{s}^{(1)}_r \circ \ldots \circ \ma{s}^{(P)}_r \in \bb{C}^{N_1 \times \cdots \times N_P},
\end{equation}
where $R$ is the number of components and $\ma{s}^{(p)}_r \in \bb{C}^{N_p \times r}$ is the $r$-th column of the $p$-th factor matrix $\ma{S}^{(p)} = [ \ma{s}^{(1)}_1, \ldots , \ma{s}^{(p)}_R] \in \bb{C}^{N_p \times R}$, for $p = \{1,\ldots,P\}$. 

Note that,  applying (\ref{p4}) into (\ref{eq:fac_s}), 
(\ref{eq:tenS_parafac}) is a straight-forward generalization where we have $R$ components, and the approximation comes from the fact \textcolor{black}{that an independent phase-shift is fitted } as a combination of $P \times R$ sets of phase-shifts, thus an approximation error is expected. However, as it will be explained in Section~\ref{Sec:Simulation_Results}, for scenarios with moderate/strong \ac{LOS} components (approximated rank-one channels) \textcolor{black}{the effect of the} fitting error  on the \ac{SE} performance is negligible.

The RX estimates the factor components solving the following problem
\begin{align}
  \label{eq:parafac_min}  \underset{\text{for} \hspace{0.1cm} r = 1,\ldots, R}{\left[ \ma{\hat{s}}^{(1)}_r, \ldots ,\ma{\hat{s}}^{(P)}_r \right]} = \underset{\ma{s}^{(1)}_r, \ldots ,\ma{s}^{(P)}_r}{\text{argmin}} \left|\left| \ten{S} - \sumx{r=1}{R} \ma{s}^{(1)}_r \circ \ldots \circ \ma{s}^{(P)}_r \right|\right|^2_\text{F},
\end{align}
where $\ma{s}^{(p)}_r \in \bb{C}^{N_p \times 1}$ is the $p$-th factor component. Let us define $\ma{S}^{(p)} = \left[\ma{s}^{(p)}_1, \ldots , \ma{s}^{(p)}_R \right] \in \bb{C}^{N_p \times R}$ as the $p$-th factor matrix, for $p = \{1, \ldots,P \}$. From  (\ref{eq:nmode_parafac}), the $p$-mode unfolding of $\ten{S}$, defined as $\nmode{S}{p} \in \bb{C}^{N_p \times N_1\cdots N_{p-1}N_{p+1} \cdots N_P }$, is given as 
\begin{equation}
 \label{eq:nmode_irs_parafac}   \nmode{S}{p} \approx \ma{S}^{(p)}\left(\ma{S}^{(P)} \diamond \ldots \diamond \ma{S}^{(p+1)} \diamond \ma{S}^{(p-1)} \diamond \ldots \diamond \ma{S}^{(1)} \right)^{\text{T}}.
\end{equation}
  \begin{algorithm}[t]
	\begin{algorithmic}[1]
		\caption{PARAFAC-IRS ALS}
		\label{Alg:ALS_PARAFAC_IRS}
		\State \textbf{Inputs}: Tensor $\ten{S} \in \bb{C}^{N_1 \times \cdots \times N_P}$, the number of components $R$, and maximum number of iterations $I$.
		\State \textcolor{black}{Randomly} initialize the factors $\ma{\hat{S}}^{(2)}_0$, $\ldots$, $\ma{\hat{S}}^{(P)}_0$. Iteration ${i} = 0$.
		\State Define a maximum number of iteration $I$.  
\For{$i = 1:I$}		
		\For{$p =1:P$}		
			\State Compute an estimate \textcolor{black}{of} the $p$-th factor  $\ma{S}^{(p)}_{\text{i}}$ as
			\begin{align*}
			\ma{\hat{S}}^{(p)}_{i}\hspace{-0.1cm} =  \hspace{-0.1cm}\nmode{S}{p} \hspace{-0.1cm}\left(\hspace{-0.1cm} \left(\ma{\hat{S}}^{(P)}_{i-1} \hspace{-0.06cm}\diamond \ldots \diamond \hspace{-0.1cm} \ma{\hat{S}}^{(p+1)}_{i-1} \hspace{-0.06cm}\diamond \ma{\hat{S}}^{(p-1)}_{i-1} \hspace{-0.06cm}\diamond \ldots \diamond \ma{\hat{S}}^{(1)}_{i-1}  \right)^{\text{T}} \right)^{\hspace{-0.06cm}+}
\end{align*}
	\For{$r =1:R$}
	\State Normalize the $r$-th column of $\ma{\hat{S}}^{(p)}_{(i)}$, defined as $\ma{s}^{(p)}_{r,(i)}$, and store its norm as the $r$-th  element of the vector $\ma{\lambda}^{(p)} \in \bb{R}^{R \times 1}$
		\begin{align*}
	\lambda^{(p)}_r &= \left|\left| \ma{\hat{s}}^{(p)}_{r,(i)}\right|\right|_2, \,\,
	&\ma{\hat{s}}^{(p)}_{r,(i)} = \frac{\ma{\hat{s}}^{(p)}_{r,(i)} }{ \lambda^{(p)}_r}.
	\end{align*}
	\EndFor							
	\EndFor
	\State Define the weighting vector $\ma{\lambda} = \ma{\lambda}^{(1)} \odot \cdots \odot \ma{\lambda}^{(P)} \in \bb{C}^{R \times 1}$.
	\State $i = i +1$
		\EndFor
		\State Return $\ma{\hat{S}}^{(1)}$, $\ldots$, $\ma{\hat{S}}^{(P)}$ and $\ma{\lambda}$.
	\end{algorithmic}
\end{algorithm}
To solve the problem in (\ref{eq:parafac_min}), the RX can use the \ac{ALS} algorithm \cite{kolda}, described in Algorithm \ref{Alg:ALS_PARAFAC_IRS}. Basically, the ALS algorithm contains \textcolor{black}{$I$ iterations, where, in each iteration, $P$ LS problems are solved. The $p$-th LS problem is defined as }
\begin{equation}
 \label{eq:parafac_min_S_P}\ma{\hat{S}}^{(p)} = \underset{\ma{S}^{(p)}}{\text{argmin}} \left|\left| \begin{split}  \ma{\hat{S}}^{(p)} - \ma{S}^{(p)}\big{(}\ma{S}^{(P)} &\diamond \ldots \diamond \ma{S}^{(p+1)} \diamond  \\
\ma{S}^{(p-1)}&\diamond \ldots \diamond \ma{S}^{(1)} \big{)}^{\text{T}}  \end{split}\right|\right|^2_\text{F}, 
\end{equation}
where its solution is given by
\begin{equation}
 \label{eq:parafac_min_S_p_sol} \ma{\hat{S}}^{(p)} = \nmode{S}{p} \left(\left( \ma{S}^{(P)} \diamond \ldots \diamond \ma{S}^{(p+1)} \diamond \ma{S}^{(p-1)} \diamond \ldots \diamond \ma{S}^{(1)} \right)^{\text{T}}\right)^{+}.
\end{equation}
In the first iteration, the first step is to estimate $\nmode{S}{1}$ based on (\ref{eq:parafac_min_S_p_sol}), for $p=1$. Then, its $R$ columns are normalized to unit norm and stored in as elements of  the vector $\ma{\lambda}^{(1)} \in \bb{R}^{R \times 1}$. After the normalization, the estimated $\nmode{\hat{S}}{1}$ is plugged in the LS  solution (\ref{eq:parafac_min_S_p_sol}) for $p=2$. Likewise, the columns of the estimated factor $\nmode{\hat{S}}{2}$ are normalized and stored in a vector defined  $\ma{\lambda}^{(2)} \in \bb{R}^{R \times 1}$, and then, the normalized estimations $\nmode{\hat{S}}{1}$ and $\nmode{\hat{S}}{2}$ are plugged \textcolor{black}{into
} the LS solution (\ref{eq:parafac_min_S_p_sol}) for $p=3$. This process continues for the $P-3$ remaining LS problems. Then, we compute the weighting vector $\ma{\lambda} \in \bb{R}^{R \times 1}$ as the Hadamard product of all $P$  factors norms, i.e., $\ma{\lambda} = \ma{\lambda}^{(1)} \odot \ma{\lambda}^{(2)} \odot \cdots \odot \ma{\lambda}^{(P)} $, finalizing the first iteration of the ALS. Then, the process repeats for all $I$ iterations or until reaching a convergence threshold by checking the \ac{NMSE} of the reconstructed tensor in \textcolor{black}{a window of} consecutive iterations. The \ac{NMSE} at the $i$-th iteration is given as

\begin{align*}
\text{e}_{(i)}  =  \frac{\left|\left| \left[ \ten{S} \right]_{(1),(i)}  - \left[ \hat{\ten{S}} \right]_{(1),{(i)}}  \right|\right|^{2}_{\text{F}}}{\left|\left| \left[ \ten{S} \right]_{(1),(i)}\right|\right|^{2}_{\text{F}}},
\end{align*}
where $\left[ \hat{\ten{S}} \right]_{(1),(i)}$ is the reconstructed $1$-mode unfolding at the $i$-th ALS iteration, given by

\begin{equation}
\left[ \hat{\ten{S}} \right]_{(1),(i)} =  \hat{\ma{S}}^{(1)} \text{diag}\left( \ma{\lambda}\right)\left( \hat{\ma{S}}^{(P)} \diamond \ldots \diamond \hat{\ma{S}}^{(2)} \right)^{\text{T}}.
\end{equation}

If $ |\text{e}_{(i)} - \text{e}_{(i-1)}| \leq \epsilon$,   where $\epsilon$ is a pre-defined threshold, the algorithm stops \cite{kolda}. \textcolor{black}{In this paper, we consider $\epsilon = 10^{-6}$}.


After the ALS algorithm, the phase-shifts of each factor and the weighting vector $\ma{\lambda}$ are quantized to be conveyed to the IRS controller. In this case, the feedback duration is given by

\begin{equation}
    \label{eq:parafac_tf} T_{\text{F}}^{(\text{PARAFAC})} =  \frac{ \text{T}_{\text{PR}} + R \sumx{p=1}{P} N_p \cdot b^{(p)}_{\text{F}} + (R-1) \cdot b^{(\text{w})}_{\text{F}} }{B_{\text{F}} \text{log}\left( 1+ \frac{p_{\text{F}}|{g}_{\text{F}}|^2}{B_\text{F} N_0} \right)},
\end{equation}
where $ \text{T}_{\text{PR}}$ is the number of bits required for a preamble of the frame, in order to inform the IRS controller the factorization parameters, such as $P$, $R$ and the quantization bits $b^{(p)}_{\text{F}}$ and $b^{(\text{w})}_{\text{F}}$, where the $b^{(p)}_{\text{F}}$ is the number of bits used for quantize the phase-shifts of the $p$-th factor, while  $b^{(\text{w})}_{\text{F}}$ is the number of bits for quantizing the elements of the weighting vector $\ma{\lambda}$.

\begin{figure}[!t]
	\centering\includegraphics[scale=0.65]{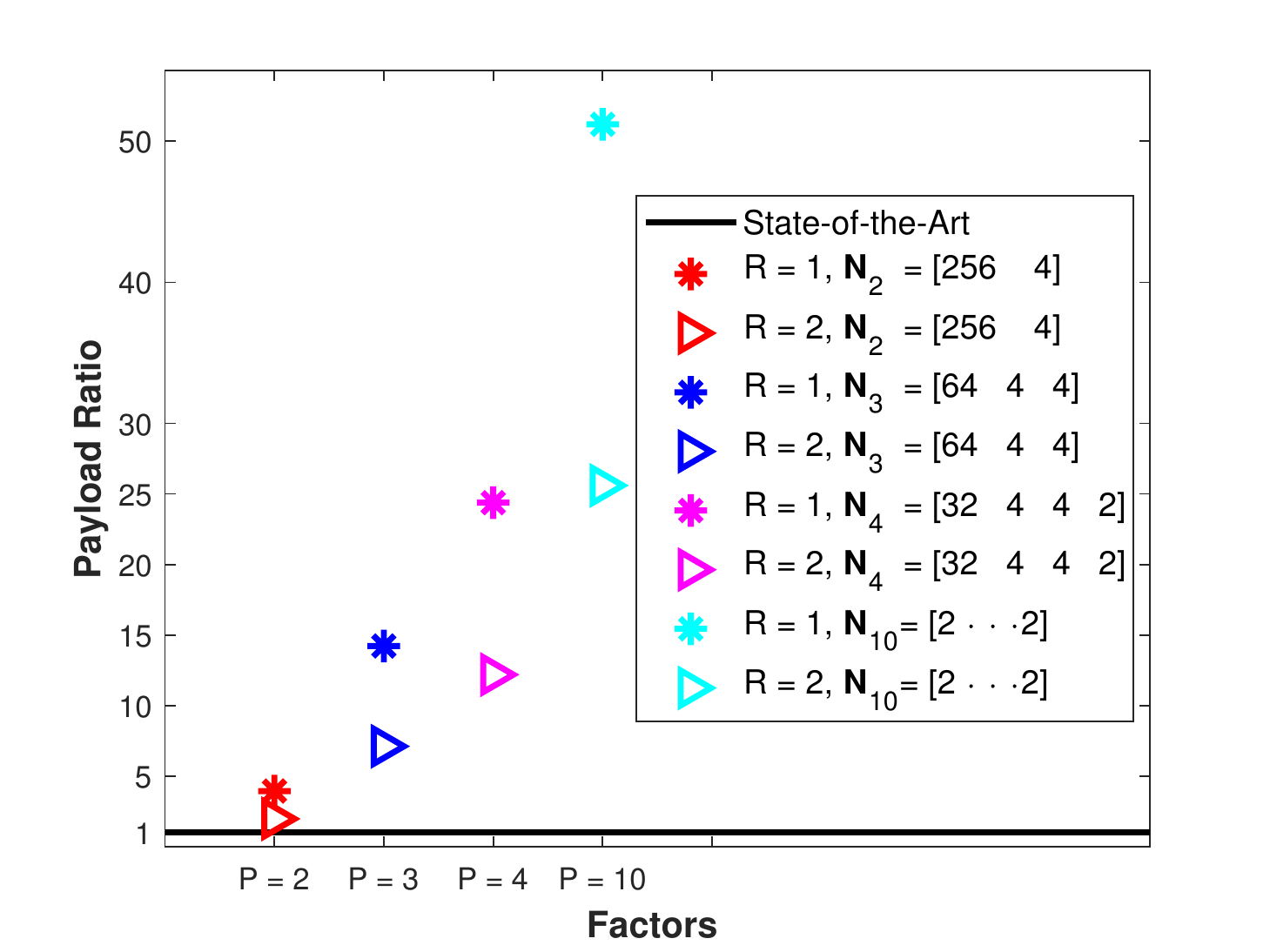}
	\caption{Feedback payload ratio for $N=1024$. }
	\label{fig:ALS_N_1024_b_3_ratio}
\end{figure}

As one example, Fig. \ref{fig:ALS_N_1024_b_3_ratio} illustrates the ratio between state-of-the-art approach, where the $N$ IRS phase-shifts are fed back, with the proposed PARAFAC-IRS approach, i.e.,  $N / (R \sumx{p=1}{P}N_P)$, not taking into account the preamble $ \text{T}_{\text{PR}}$ and the resolution. Let us define a vector $\ma{N}_{\text{P}} = \left[N_1 \ldots N_P\right] \in \bb{R}^{P \times 1}$ that contains the factor's size for a certain $P$. We can observe that, for $R=1$, the feedback duration of the proposed approach for the case of $P=2, N_1 = 256, N_2 =4$, is almost five times smaller than the state-of-the-art, while, when we increase the number of factors $P$,  the size of the factors \textcolor{black}{can be reduced}, thus decreasing the feedback duration, such that, when we have $P=10$ and $N_p = 2$, for $p=\{1,\ldots,P \}$,  the feedback overhead of the proposed approach is approximately fifty times smaller than the state-of-the art \cite{Zappone_Overhead_Aware}. As noticed, \textcolor{black}{with increasing $P$}, the feedback duration of our proposed approach decreases, but also, as it will be discussed in Section \ref{Sec:Simulation_Results}, the \ac{SE} in \ac{NLOS} scenarios. Thus, to overcome this loss the \ac{RX} can increase the number of components $R$ at the cost of a higher feedback overhead. \textcolor{black}{In this way}, the proposed overhead-aware method shows off a trade-off between \ac{SE} and feedback overhead.  

\subsection{Tucker-IRS Low-Rank Approximation}\label{Sec:Tucker_IRS}
Let us consider that the RX node opts to fit the phase-shift tensor $\ten{S}$ as a Tucker model, i.e.,
\begin{equation}
\label{eq:tenS_tucker} \ten{S}  \approx \sumx{r_1 = 1}{R_1} \ldots \sumx{r_P = 1}{R_P} \ten{G}_{r_1,\ldots,r_P} \left( \ma{s}^{(1)}_{r_1} \circ \ldots \circ \ma{s}^{(P)}_{r_P} \right) \in \bb{C}^{N_1 \times \ldots \times N_P}.
\end{equation}
where $\ten{G} \in \bb{C}^{R_1 \times \ldots \times R_P}$ is the $P$-th order core tensor and $\ma{s}^{(p)}_{r_p} \in \bb{C}^{N_p \times}$ is the $r_p$-th column of the $p$-th factor matrix $\ma{S}^{(p)} \in \bb{C}^{N_p \times R_p}$, for $p = \{1,\ldots, P \}$ and $r_p = \{1,\ldots, R_p\}$. According to (\ref{eq:tucker_mode_n}), the $p$-th mode unfolding of $\ten{S}$ is given by
\begin{equation}
    \label{eq:tucker_mode_n}\nmode{S}{p} \hspace{-0.1cm}=  \hspace{-0.1cm}\ma{S}^{(p)} \hspace{-0.07cm}\nmode{G}{p}  \hspace{-0.06cm}\left( \hspace{-0.07cm} \ma{S}^{(P)}\hspace{-0.07cm} \otimes\hspace{-0.05cm} \ldots \hspace{-0.05cm}\otimes \hspace{-0.05cm}\ma{S}^{(p+1)} \hspace{-0.05cm}\otimes \hspace{-0.05cm} \ma{S}^{(p-1)} \hspace{-0.05cm}\otimes \hspace{-0.05cm}\ldots\hspace{-0.05cm} \otimes\hspace{-0.05cm} \ma{S}^{(1)}\hspace{-0.08cm} \right)^{\text{T}}
\end{equation}

Based on the Tucker-IRS model, the RX estimates each factor matrix $\ma{S}^{(p)} \in \bb{C}^{N_p \times R}$, for $p = \{1, \ldots, P\}$, and the core tensor $\ten{G}$. This estimation procedure can be performed using, e.g., the \ac{HOSVD} algorithm  \cite{Lathauwer2000bestr1r2} \textcolor{black}{given in Algorithm \ref{Alg:HOSVD-IRS}}, \textcolor{black}{which, in this case, consists of the \ac{RX} estimating  the factors matrices by computing the \ac{SVD} of all  $P$-mode unfolding matrices of $\ten{S}$ independently.  } Defining the SVD  of $\nmode{S}{p}$ as $\ma{U}^{(p)}\ma{\Sigma}^{(p)}\ma{V}^{\text{H}}$, an estimate of $\ma{S}^{(p)}$ is given by
\begin{equation}
    \label{eq:nmode_est_hosvd} \ma{\hat{S}}^{(p)} = \ma{U}^{(p)}_{.1:R_p}  \in \bb{C}^{N_p \times R_p},
\end{equation}
which is the truncation of the left singular matrix $\ma{U}^{(p)}$ to its first $R_P$ columns, $p = \{1,\ldots, P\}$. The diagonal of the truncated singular matrix $\ma{\Sigma^{(p)}}$, defined as $ \ma{\sigma}^{(p)} = \text{diag}(\ma{\Sigma}^{(p)}_{1:R_p,1:R_p}) \in \bb{C}^{R_p \times 1}$, is stored to provide the weights to the $R_p$ components in the quantization procedure. 
Once the $P$ factor matrices are estimated, the RX obtains an estimate of the core tensor $\ten{G}$ as 

{\color{black}
\begin{equation}
    \ma{\hat{g}} = \left(\ma{\hat{S}}^{(P)} \otimes \ldots \otimes \ma{\hat{S}}^{(1)} \right)^{\text{H}} \ma{s}, \in \bb{C}^{R_1 \cdots R_P \times 1},
\end{equation}
}
where $\ma{\hat{g}} = \text{vec}\left( \hat{\ten{G}}\right)$ and $\ma{s} = \text{vec}\left(\ten{S}\right)$ are the vectorization of the core tensor and the IRS phase-shift tensor, respectively.

\textcolor{black}{The feedback duration of the Tucker-IRS model is given as}
\begin{equation}
    \label{eq:tucker_tf} T_{\text{F}}^{(\text{Tucker})} \hspace*{-0.08cm}= \hspace*{-0.05cm} \frac{ \text{T}_{\text{PR}} +  \hspace*{-0.1cm}\left(\sumx{p=1}{P} R_p N_p  b^{(p)}_{\text{F}}\right) \hspace*{-0.08cm}+\hspace*{-0.08cm} \prod\limits_{p=1}^{P} R_p +  b^{(\text{w})}_{\text{F}}\hspace*{-0.08cm} \prod\limits_{p=1}^{P}\hspace*{-0.05cm}(R_p-1)   }{B_{\text{F}} \text{log}\left( 1+ \frac{p_{\text{F}}|{g}_{\text{F}}|^2}{B_\text{F} N_0} \right)},
\end{equation}
\textcolor{black}{where $\text{T}_{\text{PR}}$ is the preamble duration that informs to the IRS controller the chosen \ac{LRA} model, the number of factors $P$, and the number of components $R_p$, for $p=\{1,\ldots,P\}$. The term $  \sumx{p=1}{P} R_p N_p  b^{(p)}_{\text{F}} $  represents the cost, in bits, of the conveyed phase-shifts, $\prod\limits_{p=1}^{P} R_p $ is the cost of the phase-shifts of the core tensor, and  $b^{(\text{w})}_{\text{F}}\hspace*{-0.08cm} \prod\limits_{p=1}^{P}(R_p-1)  $ is the term related to the cost of the weighting factors.}
%
%

{\color{black}
\section{Discussion on Quantization, Reconstruction and Parameter Choices}
\label{eq:quan_rec_par}
}

\subsection{Phase-shift Quantization}\label{Sec:ps_quantization}

After estimating the factors  in Algorithms~\ref{Alg:ALS_PARAFAC_IRS} or \ref{Alg:HOSVD-IRS}, the \ac{RX} quantizes the phase-shifts of each factor  with $b_F^{(p)}$ bits. Let us define  $ \ma{\tilde{a}} = \mathcal{Q}\{\ma{a},b \}$ as the quantization operation, which quantizes a phase-shift vector $\ma{a}$ with $b$ bits.  For the PARAFAC-IRS model, we have  the following quantized factors $\ma{\tilde{s}}_r= \mathcal{Q}\left\{\ma{\hat{s}}^{(p)}_r,b_\text{F}^{(p)}\right\}$ for $p=\{1,\ldots, P\}$ and $r = \{1,\ldots,R\}$. In addition, for the Tucker-IRS model, we have the following quantized  factors and core tensor
$\ma{\tilde{s}}_{r_p}= \mathcal{Q}\left\{\ma{\hat{s}}^{(p)}_{r_p},b_\text{F}^{(p)}\right\}$ and $\tilde{\ten{G}}_{r_1,\ldots,r_P}= \mathcal{Q}\left\{\hat{\ten{G}}_{r_1,\ldots,r_P},b_\text{F}^{(p)}\right\}$, for $p=\{1,\ldots, P\}$ and $r_p = \{1,\ldots,R_p\}$. For the phase-shift quantization of the $p$-th factor, we use the following codebook 

\begin{align*}
\mathcal{C}_{\phi}^{(p)} = \left\{ -\pi + \frac{2\pi }{2^{b^{(p)}_{\text{F}}} },\hspace{0.2cm} -\pi + \frac{4\pi }{2^{b^{(p)}_{\text{F}}} }, \hspace{0.2cm} \ldots, \hspace{0.2cm} \pi \right\}.
\end{align*}

\subsection{Weighting Factor Quantization}\label{Sec:wf_quantization}

For the PARAFAC-IRS model, let us define $\lambda_{\text{max}}$ as the largest element of $\ma{\lambda}$. Then, we define a new weighting vector $\ma{\lambda}^{\prime} = \ma{\lambda}/\lambda_{\text{max}} \in \bb{R}^{R \times 1}$. Since the largest element of $\ma{\lambda}^{\prime}$ is one, we do not need to quantize this element. Hence, we define a new vector $\bar{\ma{\lambda}} \in \bb{R}^{R-1 \times 1}$ that contains all elements of $\ma{\lambda}^{\prime}$, with the exception the of the largest one. Then, we quantize the weighting vector by defining $\tilde{\bar{\ma{\lambda}}} = \mathcal{Q}\left\{\bar{\ma{\lambda}}, b^{(\text{w})}_{\text{F}} \right\}$. Finally, we define $\tilde{\ma{\lambda}} \in \bb{R}^{R\times 1}$ as the quantized weighting vector by inserting in the correct position the largest element of $\ma{\lambda}^{\prime}$ (one) in $\tilde{\bar{\ma{\lambda}}} \in \bb{R}^{R-1 \times 1}$. At the end, the weighting vector quantization cost is $(R-1) b^{\text{w}}_{\text{F}}$ bits.

For the Tucker model, a similar approach is made, with the difference that there are $P$ weighting vectors sorted by their largest value due to the SVD procedure. Considering the $p$-th weighting vector $\ma{\sigma}^{(p)} \in \bb{R}^{R_p \times 1}$, we normalize it by the first element, yielding $\ma{\sigma}^{(p)\prime}= \ma{\sigma}^{(p)}/\sigma^{(p)}_1$. For the quantization, we define a vector $\bar{\ma{\sigma}}^{(p)} \in \bb{R}^{R_p -1 \times 1}$ that contains all elements of $\ma{\sigma}^{(p)\prime}$ with exception of the first one. Then, we define the quantized $p$-th weighting factor as $\tilde{\bar{\ma{\sigma}}}^{(p)} = \mathcal{Q}\left\{\bar{\ma{\sigma}}^{(p)}, b^{(\text{w})}_{\text{F}} \right\} \in \bb{R}^{R_p-1 \times 1}$. Finally, for the $p$-th quantized vector, we define the quantized vector $\tilde{\ma{\sigma}}^{(p)} = [1,  \tilde{\ma{\sigma}}^{(p)\prime}] \in \bb{R}^{R_p  \times 1}$. At the end, the quantization of the $P$ weighting factors for the Tucker model costs $b^{\text{w}}_{\text{F}} \cdot \prod\limits_{p=1}^{P} \left( R_p - 1 \right) $ bits.

For both PARAFAC and Tucker models, we define the following amplitude codebook
\begin{align}
\label{eq:amp_code} \mathcal{C}_{\text{w}} = \left\{ 0.01, \hspace{0.2cm} 0.01 + l, \hspace{0.2cm} 0.01 + 2 l, \hspace{0.2cm} \ldots, \hspace{0.2cm} 1 \right\},
\end{align}
where  $l=\frac{1-0.01}{2^{b^{(\text{w})}_{\text{F}}}-1}$ is the pre-defined step. For simplicity, the values of the amplitudes in (\ref{eq:amp_code}) are rounded to the second decimal point.

\subsection{IRS Phase-shift Vector Reconstruction}\label{Sec:irs_recon}
After quantization, the \ac{RX} conveys the factors to the IRS controller. Then, \textcolor{black}{the phase-shift vector} is reconstructed as
\begin{align}
\label{eq:recon} \ma{s} = e^{j \angle \ma{\hat{s}}} \in \bb{C}^{N \times 1},
\end{align}
where $\ma{\hat{s}}$ is given by
\begin{align}
\label{eq:recon_parafac}\ma{\hat{s}} = \sum\limits_{r=1}^R  \tilde{\lambda}_r \left( \ma{\tilde{s}}^{(P)}_r \otimes \ldots \otimes \ma{\tilde{s}}^{(1)}_r\right),
\end{align}
for the PARAFAC-IRS model, while for the Tucker-IRS model, $\ma{\hat{s}}$ is factorized as
\begin{align}
\label{eq:recon_tucker_t}\ma{\hat{s}}= \hspace*{-0.12cm} \sumx{r_1 = 1}{R_1} \hspace*{-0.1cm}\ldots \hspace*{-0.1cm}\sumx{r_P = 1}{R_P} \tilde{\ten{G}}_{r_1,\ldots,r_P} \left( \tilde{\sigma}^{(P)}_{r_P}\ma{\tilde{s}}^{(P)}_{r_P}\right) \otimes \ldots \otimes \left(\tilde{\sigma}^{(1)}_{r_1}\ma{\tilde{s}}^{(1)}_{r_P} \right).
\end{align}

\begin{algorithm}[t]
	\begin{algorithmic}[1]
		\caption{Tucker-IRS HOSVD}
		\label{Alg:HOSVD-IRS}
		\State \textbf{Inputs}: Tensor $\ten{S}$, the number  of components $R_p$, for $p = \{1,\ldots, P\}$.
		\For{$p =1:P$}		
			\State Compute the SVD of the $p$-mode unfolding of $\ten{S}$ as  
			\begin{align*}
			\nmode{S}{p} = \ma{U}^{(p)}\ma{\Sigma}^{(p)} \ma{V}^{(p)\text{H}}.
\end{align*}
\State Store the diagonal of the truncated  singular matrix defined as $\ma{\sigma}^{(p)} = \text{diag}\left(\ma{\Sigma}^{(p)}_{1:R_p,1:R_p}\right) \in \bb{C}^{R_p \times 1}$.
			\State Set an estimation of $\ma{S}^{(p)}$ by truncating the left singular matrix \textcolor{black}{to its first} $R$ columns
			\begin{align*}
			\ma{\hat{S}}^{(p)} = \ma{U}^{(p)}_{.1:R_p}.
			\end{align*}						
	
		\EndFor
		\State Compute an \textcolor{black}{estimate} of the core tensor $\ma{g} = \text{vec} \left( \ten{G}\right)$ as 
\begin{align*}
\text{vec}\left( \hat{\ma{g}}\right) = \left(\ma{\hat{S}}^{(P)\text{H}} \otimes \ldots \otimes \ma{\hat{S}}^{(1)\text{H}}  \right) \text{vec}\left(\ten{S} \right).
\end{align*}
\State Define $\hat{\ten{G}} = \ten{T} \{ \text{vec}\left( \hat{\ma{g}}\right) \}$.
		\State Return $\ma{\hat{S}}^{(1)}, \ldots, \ma{\hat{S}}^{(P)}$ and $\hat{\ten{G}}$.
	
	\end{algorithmic}
\end{algorithm}

\subsection{On the Effect of the Factorization Parameters}\label{Sec:Parameters}
In this section, we discuss the choice of the factorization parameters and the system performance implications.

\begin{itemize}
\item \textbf{Number of factors} $P$: This parameter defines the total number of factors used in the \ac{LRA}.  Its minimum value for the proposed factorization is $P=2$, i.e., $P=1$ means that no factorization is employed, its maximum value is $\text{log}_2(N)$, for the case \textcolor{black}{where} all the factors have size $N_p=2$ for $ p =\{1,\ldots,P\}$. By increasing the value of $P$, \textcolor{black}{the number of factors of increases}, allowing to reduce the size of the factor components $N_p$. Consequently, increasing $P$ reduces the phase-shift feedback overhead. Nevertheless, by selecting \textcolor{black}{the minimum value of $P$}, the size of each factor component \textcolor{black}{increases, which increases the spectral efficiency at the cost of a higher feedback overhead.}

\item \textbf{Number of components}: For the PARAFAC model, we have $R$ components, while for the Tucker model we have $P \cdot \sumx{p=1}{P}R_p$ components. For both models, the number of components is a performance indicator since when increases,  the approximation error of the \ac{LRA} in (\ref{eq:tenS_parafac}) and (\ref{eq:tenS_tucker}) decreases. The RX selects its value based on the estimated channels.  For example, if the channels \textcolor{black}{have low-rank, or in the presence of a moderate/strong \ac{LOS} component, the RX} may choose $R=1$. Also, $R=1$ (PARAFAC model) or $R_p = 1$ (for the Tucker model) are the choices that minimizes the feedback overhead. On the other hand, by increasing $R$ (or $R_p$), the \ac{SE} increases at the cost of a higher feedback load.


\item \textbf{Size of factor components} $N_p$: The size of the factor components indicates the total number of independent phase-shifts in the proposed solution, which it also affects the performance.  For example, for $N=256$, $P=2$ and $R=1$ for the PARAFAC-IRS model, two possible configurations are $(N_1=128$, $N_2=2)$ and $(N_1=N_2=16)$. For the first choice, the system has more independent phase-shifts ($130$), thus a higher \ac{SE}. However, its feedback overhead is higher than that of the second configuration that requires only $32$ phase-shifts to be reported in the feedback channel. 

 \end{itemize}
 \begin{figure*}[!t]
\centering
\begin{subfigure}{.5\linewidth}
  \centering
  \includegraphics[scale=0.6]{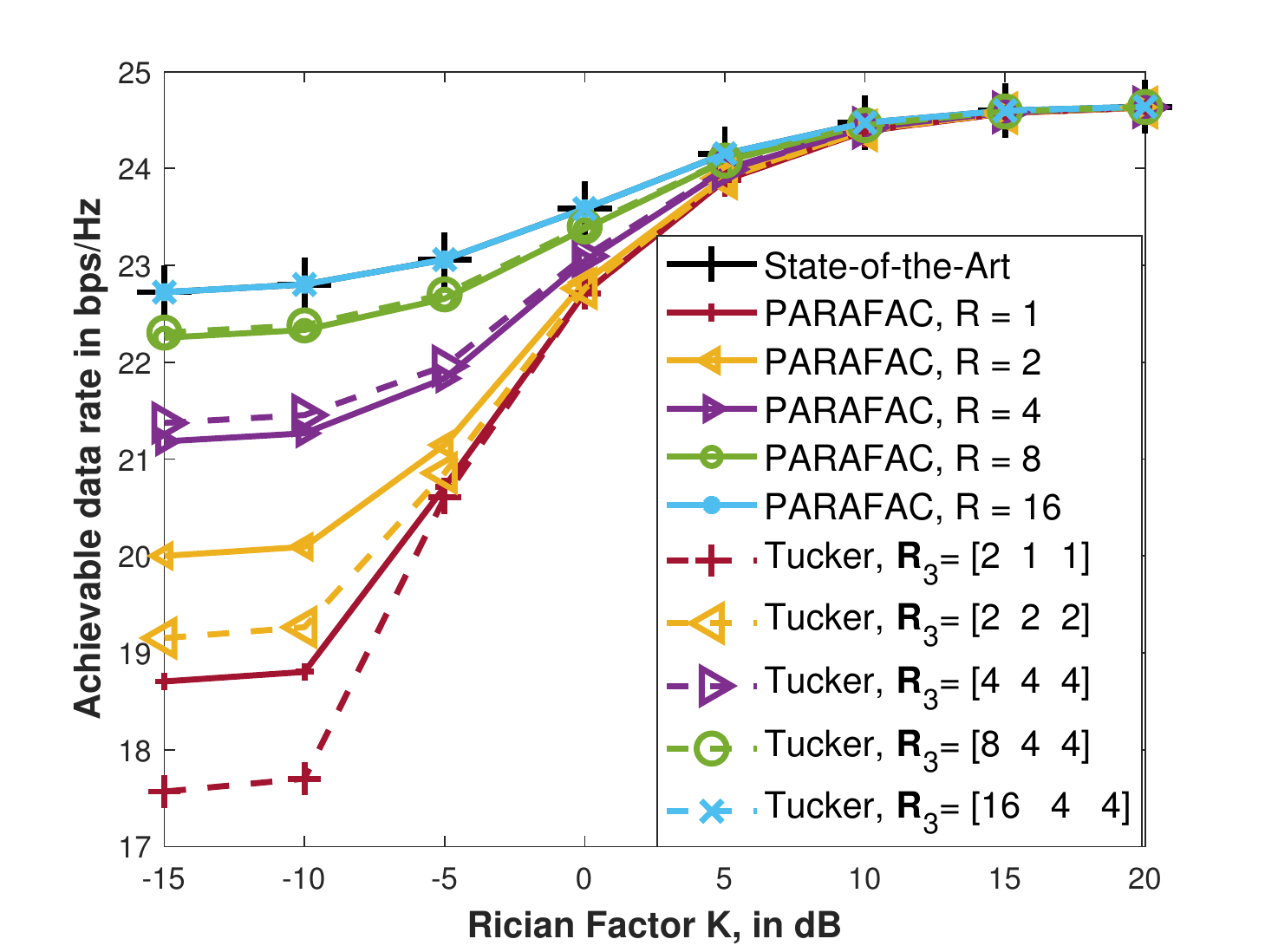}
  \caption{ No quantization and feedback.}
  \label{fig:sub1}
\end{subfigure}%
\begin{subfigure}{.5\linewidth}
  \centering
 \includegraphics[scale=0.6]{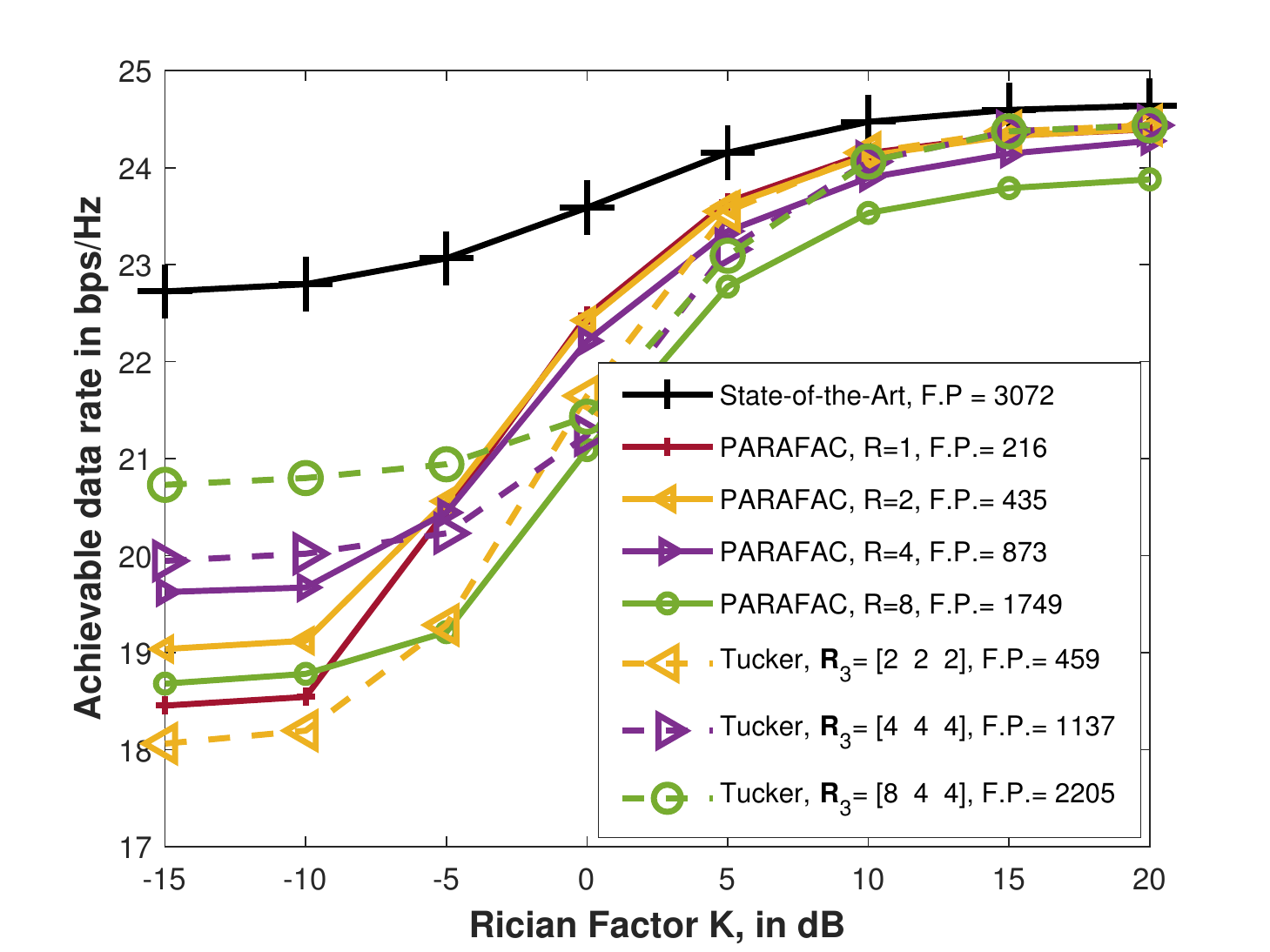}
  \caption{Comparison of the feedback payload (F.P.) in bits.}
  \label{fig:sub2}
\end{subfigure}
\caption{Comparison between the state-of-the-art \cite{Zappone_Overhead_Aware}, PARAFAC-IRS and Tucker-IRS models with different numbers of components. $N=1024$, $P=3$, with $N_1 =64$, $N_2=4$. }
\label{fig:ALS_vs_HOSVD}
\end{figure*}
 
\subsection{On the Effect of the Phase-shift and Weighting Factor  Quantization}\label{Sec:ps_wf_effect}

After the factorization step, the phase-shifts of the factor matrices $\ma{S}^{(p)}$ are quantized before being conveyed to the IRS controller. From the fact that the proposed method factorizes the IRS phase-shift vector into $P$ smaller factors, we can select  different numbers of bits for the quantization of each factor, unlike the conventional IRS-assisted systems, where the RX (or TX) conveys the $N$ phase-shifts with the same quantization resolution of $b_\text{F}$ bits. {\color{black} The proposed method allows the system to adapt the phase-shift resolution of the factors to the available control link capacity, i.e., for each factor we may have a different resolution of $b^{(p)}_{\text{F}}$ in bits, $p= \{1,\ldots, P\}$, providing more flexibility to the system design. Regarding the weighting factors, they play a more important role when the number of components $R>1$ (in the PARAFAC model), or $R_p > 1$, $p= \{1,\ldots, P\}$ (in the Tucker model), since they control the importance of the rank-one components in each model.}

\section{Simulation Results}\label{Sec:Simulation_Results}

In this section, we evaluate the performance of the proposed IRS phase-shift overhead-aware feedback model in terms of feedback duration, achievable data rate, \ac{SE} and \ac{EE}. The channels in (\ref{eq:sig}) are modeled as 
\begin{align}
\label{eq:channel_H} \ma{H} &= \sqrt{\alpha_H \frac{K_H}{K_H+1}} \ma{H}_{\text{LOS}}  +  \sqrt{\alpha_H\frac{1}{K_H+1}} \ma{H}_{\text{NLOS}}, \\
\label{eq:channel_G}\ma{G} &= \sqrt{\alpha_G \frac{K_G}{K_G+1}} \ma{G}_{\text{LOS}}  +  \sqrt{\alpha_G\frac{1}{K_G+1}} \ma{G}_{\text{NLOS}}, 
\end{align}
where $\alpha_H$ and $\alpha_G$ are the path-loss components of the \ac{TX}-IRS and IRS-\ac{RX} links, respectively. The matrices $K_H$ and $K_G$ are the Rician factors associated with $\ma{H}$ and $\ma{G}$, respectively. $\ma{H}_{\text{LOS}}$, $\ma{G}_{\text{LOS}}$ follow a geometric-based channel model, while the entries of $\ma{H}_{\text{NLOS}}$, $\ma{G}_{\text{NLOS}}$ are modeled as circularly symmetric complex Gaussian random variables, with zero mean an unit variance, i.e., $\ma{H}_{\text{NLOS}} \sim \mathcal{CN}(0,\ma{I}_{M_T})$ and $\ma{G}_{\text{NLOS}} \sim \mathcal{CN}(0,\ma{I}_{M_R})$. More details of (\ref{eq:channel_H}) and (\ref{eq:channel_G}) are given in Appendix \ref{App:channel_model}. 

For a fair comparison between the state-of-the-art \cite{Zappone_Overhead_Aware} and the proposed PARAFAC-IRS and Tucker-IRS models, we optimize the precoder ($\ma{q}$), combiner ($\ma{w}$), and the IRS phase-shifts ($\ma{s}$) using the upper-bound solution of \cite{Zappone_Overhead_Aware}. In this case, they are given as
\begin{align*}
\ma{w} = \ma{U}^{(\text{G})}_{.1}, \hspace{0.1cm} 
\ma{q} =  \ma{V}^{(\text{H})}_{.1}, \hspace{0.1cm}
\ma{s}^{(\text{opt})}_n = e^{-\angle \left(\ma{V}^{ \text{G})}_{n,1} \cdot \ma{U}^{(\text{H})}_{n,1} \right)},
\end{align*} 
with  $n = \{1, \ldots, N\}$, and $\ma{U}^{(\text{G})}_{.1} \in \bb{C}^{M_R \times 1}$, $\ma{V}^{ (\text{G})}_{.1} \in \bb{C}^{N \times 1}$ are the dominant left and right singular vectors of $\ma{G}$, while $\ma{U}^{(\text{H})}_{.1} \in \bb{C}^{N \times 1}$, $\ma{V}^{( \text{H})}_{.1} \in \bb{C}^{M_T \times 1}$ are the dominant left and right singular vectors of $\ma{H}$. 

\begin{figure}[!t]
	\centering\includegraphics[scale=0.6]{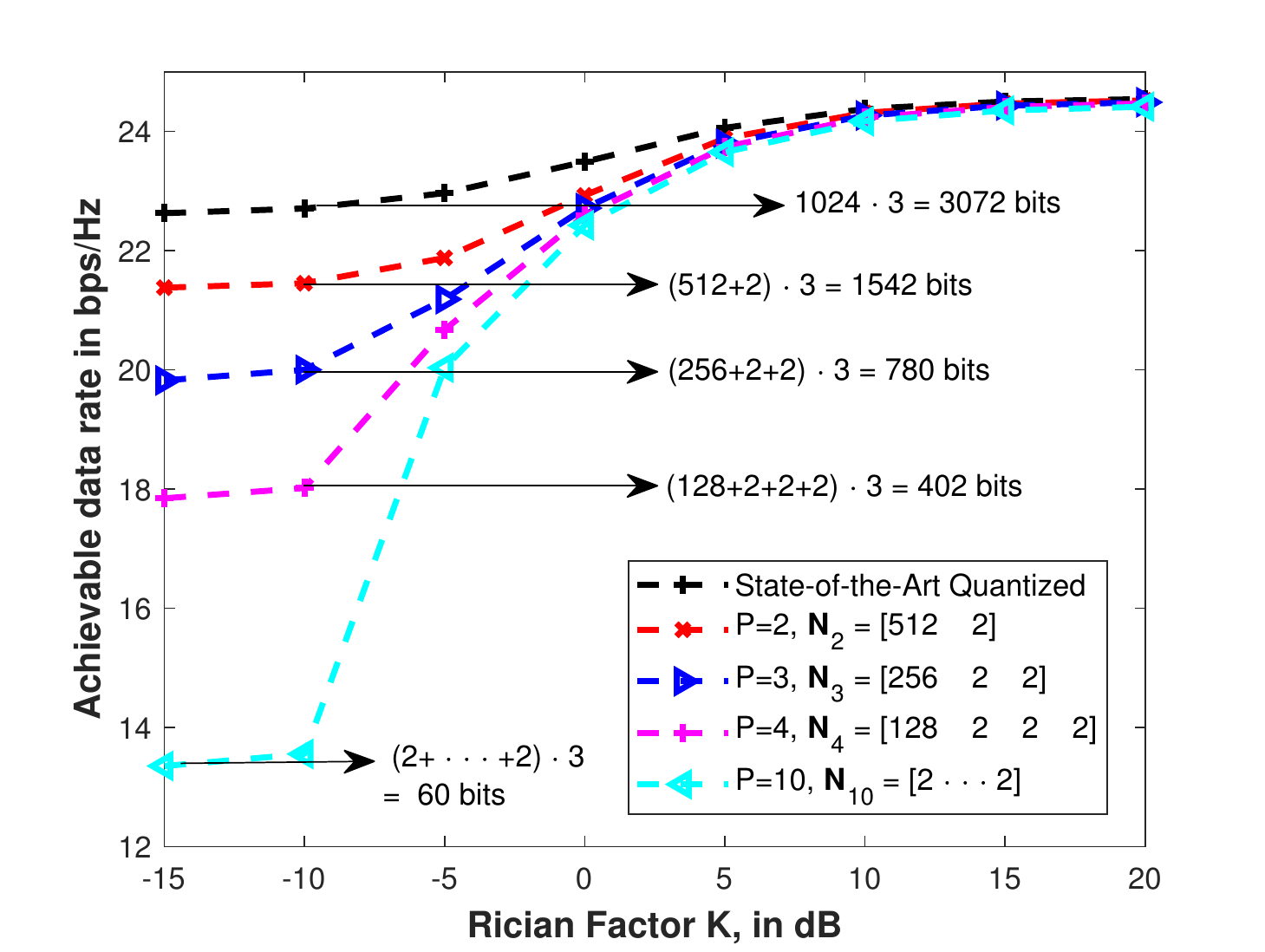}
	\caption{For an IRS with $N=1024$, \ac{TX} and \ac{RX} with $M_R=M_T=2$ and  $ b^{(p)}_\text{F} = b_{\text{F}}= 3$ bits, for the IRS phase-shift quantization resolution,  for $p = \{1, \ldots,P\}$.}
	\label{fig:varying_P}
\end{figure}

\begin{figure}[!t]
	\centering\includegraphics[scale=0.65]{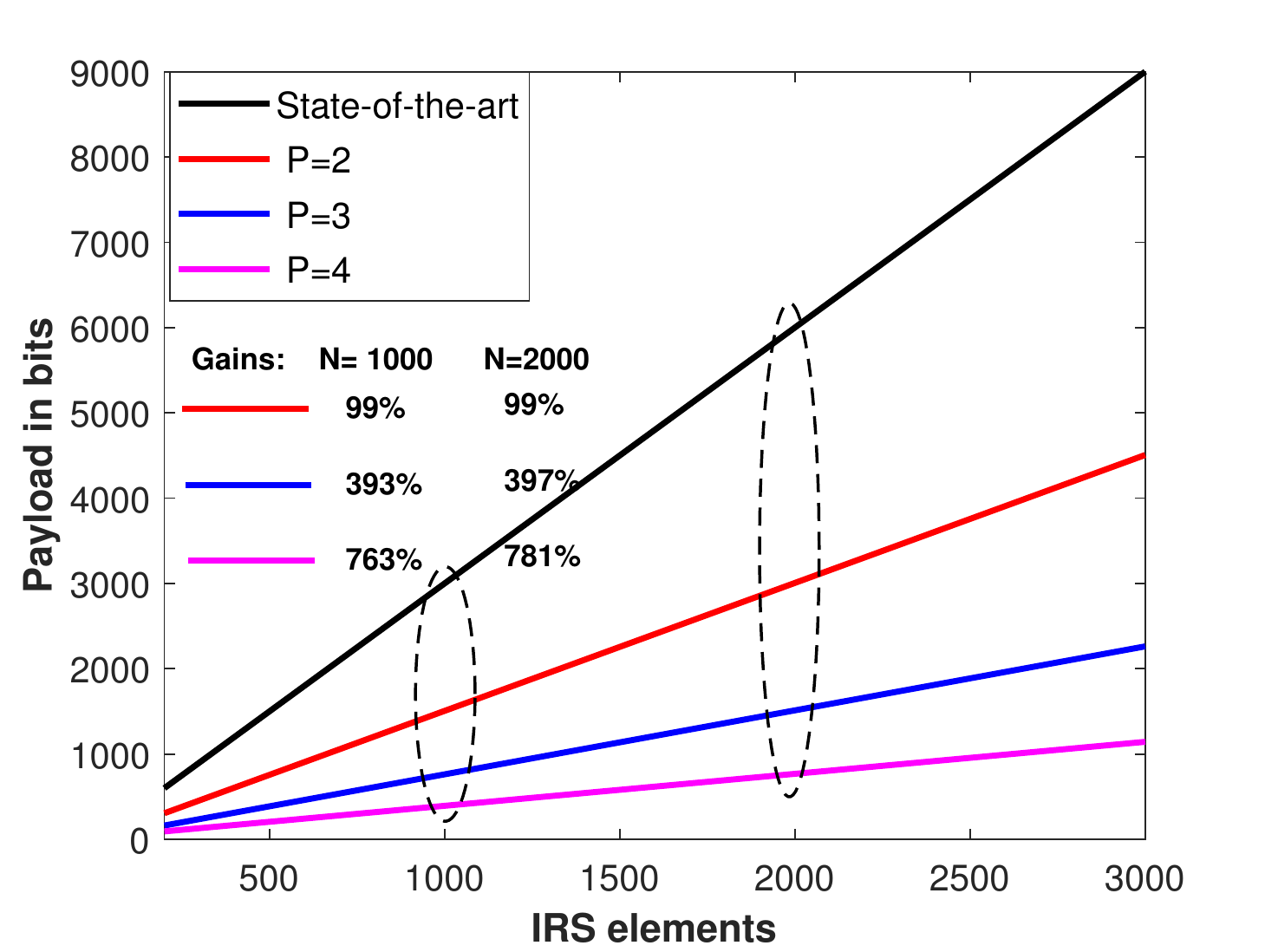}
	\caption{Feedback payload for the PARAFAC-IRS model with $R=1$, varying the number of IRS elements.}
	\label{fig:varying_N}
\end{figure}

\textcolor{black}{In Figs. \ref{fig:ALS_vs_HOSVD}-\ref{fig:varying_N}, we set $\alpha_H = \alpha_G = 1$, and consider $K_H = K_G = K$  to simplify the presentation of the figures. However, we have tested the results for a broad range of channel models and parameter settings and observed the same qualitative conclusions as those presented.}

\subsection{PARAFAC-IRS \textit{vs} Tucker-IRS}
As a first experiment, we compare, in terms of achievable data rate, the two proposed strategies, PARAFAC-IRS and Tucker-IRS, with the state-of-the-art method \cite{Zappone_Overhead_Aware}, where the IRS phase-shift vector is not factorized. The achievable data rate is given by
\begin{align}
\label{eq:ADR} \text{log}_2 \left( 1 + \frac{|\ma{w}^{\text{H}}\ma{G}\ma{S}\ma{H}\ma{q}|^2  }{\sigma^2_b}\right) \hspace{0.15cm},  \text{in bits/s/Hz},
\end{align}
where $\ma{S} = \text{diag}(\ma{s}^{(\text{opt})}) \in \bb{C}^{N \times N}$ is the diagonal matrix containing the optimum IRS phase-shifts, which are given in (\ref{eq:recon_parafac}) for the PARAFAC-IRS model and in (\ref{eq:recon_tucker_t}) for the Tucker model.

In Fig. \ref{fig:ALS_vs_HOSVD}, we assume $P=3$ for the proposed IRS factorization models, with $N_1=64$, and $N_2=N_3 =4$. As expected, in this scenario the state-of-the-art solution \cite{Zappone_Overhead_Aware}, provides the performance upper bound, since no factorization is applied.  

In Fig. \ref{fig:ALS_vs_HOSVD} (a), we compare the models in an ideal scenario with continuous phase-shift and continuous values for the weighting factors. \textcolor{black}{For simplicity}, let us define for the Tucker model, the vector $\ma{R}_{\text{P}} = \left[R_1, R_2, \ldots R_P, \right] \in \bb{R}^{P \times 1}$ that contains the number of components for each factor for a certain $P$ (with $P=3$ in this case). As expected, when the number of components $R$ or $\ma{R}_{(\text{3})}$ increases, the achievable data rate also increases, and we can observe that for the PARAFAC-IRS model with $R=16$ and for the Tucker-IRS model with $\ma{R}_3 =[16,4,4]$, the proposed models achieves the optimum performance of the benchmark method \cite{Zappone_Overhead_Aware}. 

In practice, both the phase-shift and the weighting factors have to be quantized, as illustrated in Fig. \ref{fig:ALS_vs_HOSVD} (b), there is an optimal point for the PARAFAC ($R=4$, for this case) since when $R>4$ the performance degrades \textcolor{black}{due to overfitting}. For the Tucker model, when the number of \textcolor{black}{components} of $\ma{R}_{3}$ increases, the performance in the NLOS region ($K < -5$ dB) also improves at the cost of a higher feedback \textcolor{black}{overhead}. Note that, for the moderate/strong LOS scenario ($K > 5$ dB), the number of components for both models does not give a \textcolor{black}{noticeable} performance enhancement. \textcolor{black}{In this way}, a proper model for a NLOS scenario would be the Tucker one, while PARAFAC is preferable in moderate/strong LOS cases, since it leads to the best performance with the lowest feedback cost, which can be explained by the fact the the channel matrices have low rank and the contributions of the components compared to the largest one are negligible. 



\textcolor{black}{In the following, we consider the PARAFAC-IRS model, due its simplicity and lower phase-shift and weight feedback cost. However, we have tested the results for the cases with Tucker method and observed the same qualitative conclusions as those presented.}

\begin{figure}[!t]
	\centering\includegraphics[scale=0.65]{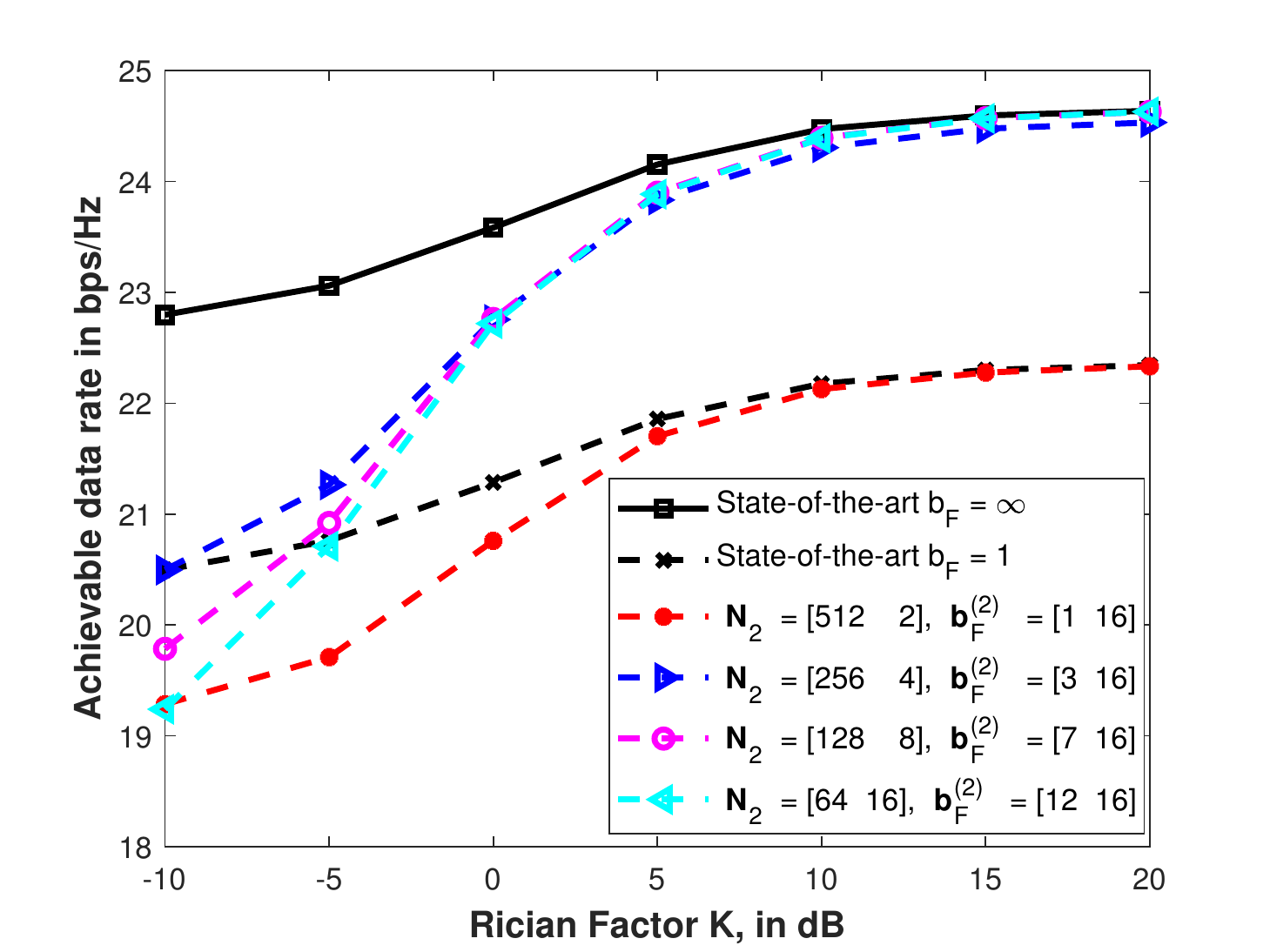}
	\caption{Performance of the PARAFAC-IRS method by varying the resolution $b^{(p)}_{\text{F}}$ per factor, for fixed control link of $1024$ bits.}
	\label{fig:fix_control_P_2}
\end{figure}

\subsection{On the Effect of the Number of Factors $P$}
In Fig. \ref{fig:varying_P}, we compare the achievable data rate of the PARAFAC-IRS model with $R=1$, by varying the number of factors. We can observe that, for the NLOS region ($K <-5$ dB), increasing $P$ leads to a degradation on the performance. This is due to the fact that, for a larger $P$, we have less independents phase-shifts. For example, for $P=10$, the phase-shifts of the IRS elements are given by the sum of $10$ factorized phase-shifts. However,  when the Rician factor $K$ increases,  the performance gap between our proposed model and the state-of-the-art \cite{Zappone_Overhead_Aware} reduces. This is explained by the fact that, the IRS phase-shift optimization is based on the channel estimation, thus when $K$ increases, the LOS components become stronger, and we have a better approximation of the PARAFAC-IRS model for $R=1$. \textcolor{black}{In terms of feedback overhead, when $P=2$ and for the $K<-5$ dB region, our proposed method has a data rate loss of approximately $1$bps/Hz. However, the feedback overhead is $50\%$ less than that of the benchmark solution \cite{Zappone_Overhead_Aware}. 
On the other hand, when the scenario changes to $K> 5$ dB, the proper parameter choice \textcolor{black}{is} $P=10$, since this configuration has a negligible performance loss compared to the state-of-the-art one, while having a lower feedback cost compared to the other proposed configurations ($P=2,3,4$).}

Physically, the results illustrated on Fig. \ref{fig:varying_P}  can be interpreted as a performance adaptation in the NLOS case, i.e., the \ac{RX} can properly choose the factorization parameters to meet a required data rate performance or feedback saving. For instance, in this example, by choosing $P=10$, the \ac{RX} can afford more often feedback than configurations with smaller values of $P$. 

For a better understanding of the merits of the proposed method, Fig.\ref{fig:varying_N} shows the feedback payload in bits by varying the number of IRS elements. \textcolor{black}{As shown, for different methods the payload increases linearly with the number of IRS elements.} For a given $P$, we may have different sets of factor sizes defined by $\ma{N}_{\text{P}} = [ N_1, \ldots, N_P]^{\text{T}} \in \bb{R}^{P \times 1}$, \textcolor{black}{where the values of $P$ are set to $P=2,3,4$}. We select the size configuration that leads to the better performance, which is the one that has the maximum possible number of independent phase-shifts. \textcolor{black}{For example, assuming $N=1000$, the size configuration is $ \ma{N}_2 = [500, 2]$  for $P=2$, $\ma{N}_3 = [250, 2, 2]$ for $P=3$, and $\ma{N}_4 = [125, 2, 2,2]$ for $P=4$.}  Thus, it becomes clear that increasing $P$ drastically reduces the feedback overhead.

\subsection{On the Effect of the Factor Quantization}

\textcolor{black}{Here, we evaluate the performance of the proposed method in a limited feedback channel, i.e., we assume that the feedback control link has a maximum capacity of $1024$ bits. In this case, traditional quantization applied to the unconstrained IRS phase shift vector (without factorization) is limited to a one bit resolution. We assume this challenging scenario to observe the performance impact of the proposed method when the resolution of the individual factors are adapted. To this end, we assume $ N \cdot b_{\text{F}} \geq \ma{N}_{\text{P}} \cdot \ma{b}^{(\text{P})\text{T}}_{\text{F}}$.  In Fig. \ref{fig:fix_control_P_2}, different sets of size configurations for $P=2$ are evaluated, with different resolutions per factor. The configuration $\ma{N}_{2} = [512,2]$ has the worst performance due to the fact that the first factor ($512$ elements) can only be quantized with $1$ bit. However, the size of the factors is reduced, the resolution per factor can be increased accordingly to meet the limited control link capacity limit. For instance, when $\ma{N}_2 = [256, 4]$ and $\ma{b}_{\text{F}}^{(p)} = [3, 16]$, the total number of bits is $256 \cdot 3 + 4 \cdot 16 = 832$. We can observe that, by increasing the resolution of the factors, the performance gets closer to that of the state-of-the-art phase shift quantization (solid curve). In particular, note that for $K> -5$ dB, our approach provides the best results. Thus, the proposed method \textcolor{black}{can not only reduce} the feedback overhead, as illustrated in Figs. \ref{fig:varying_P} and \ref{fig:varying_N}, but also it effectively provides higher data rates than traditional quantization over the unconstrained IRS phase shifts, approaching the continuous phase-shift case.
}

\begin{figure*}[!t]
\centering
\begin{subfigure}{.48\linewidth}
  \centering
\includegraphics[scale=0.6]{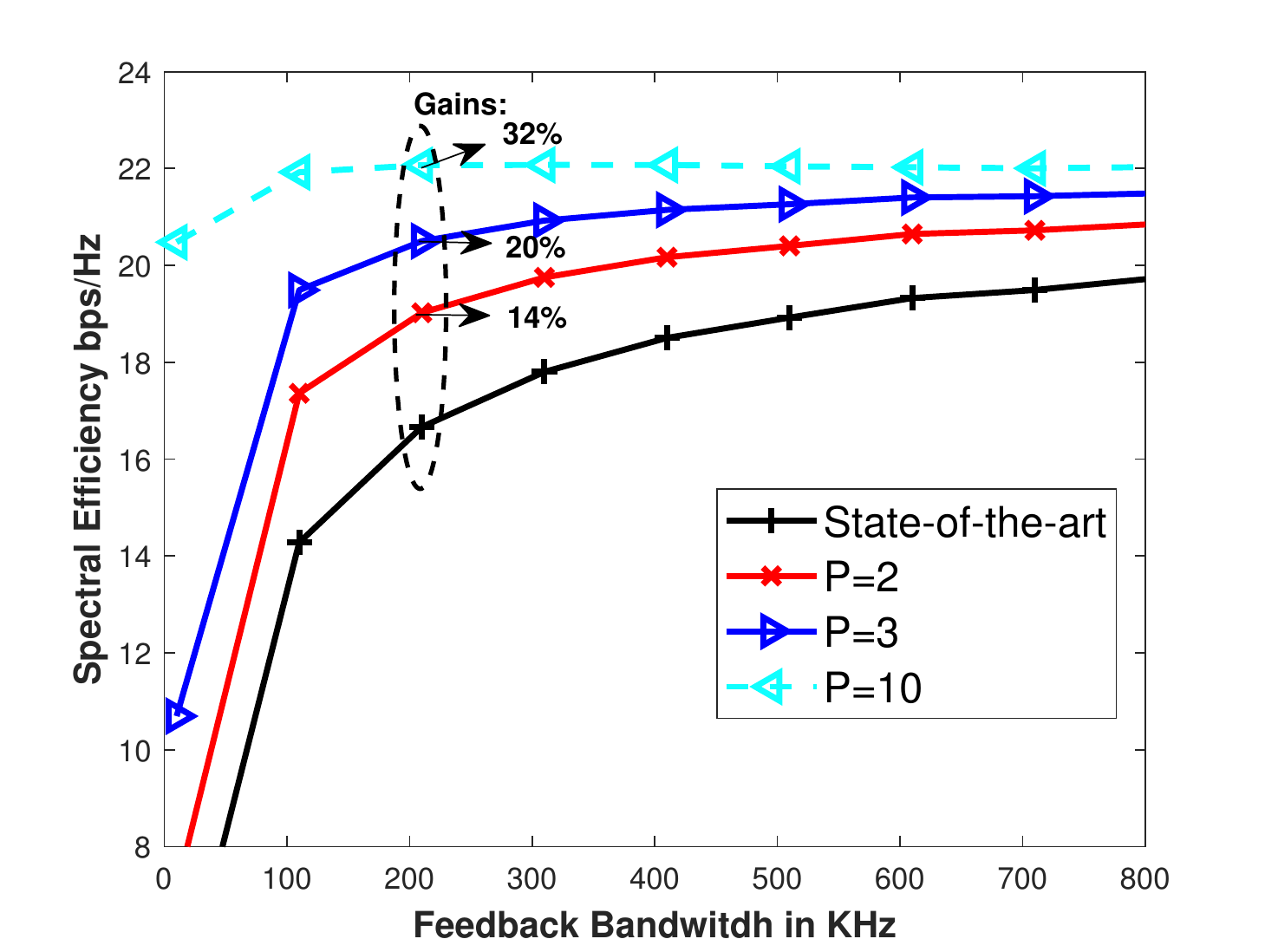}
\caption{\ac{SE} performance of the state-of-the-art and the proposed method..}
	\label{fig:SE_vary_bf_fix_M}
\end{subfigure}%
\hspace{0.2cm}
\begin{subfigure}{.48\linewidth}
  \centering
\includegraphics[scale=0.6]{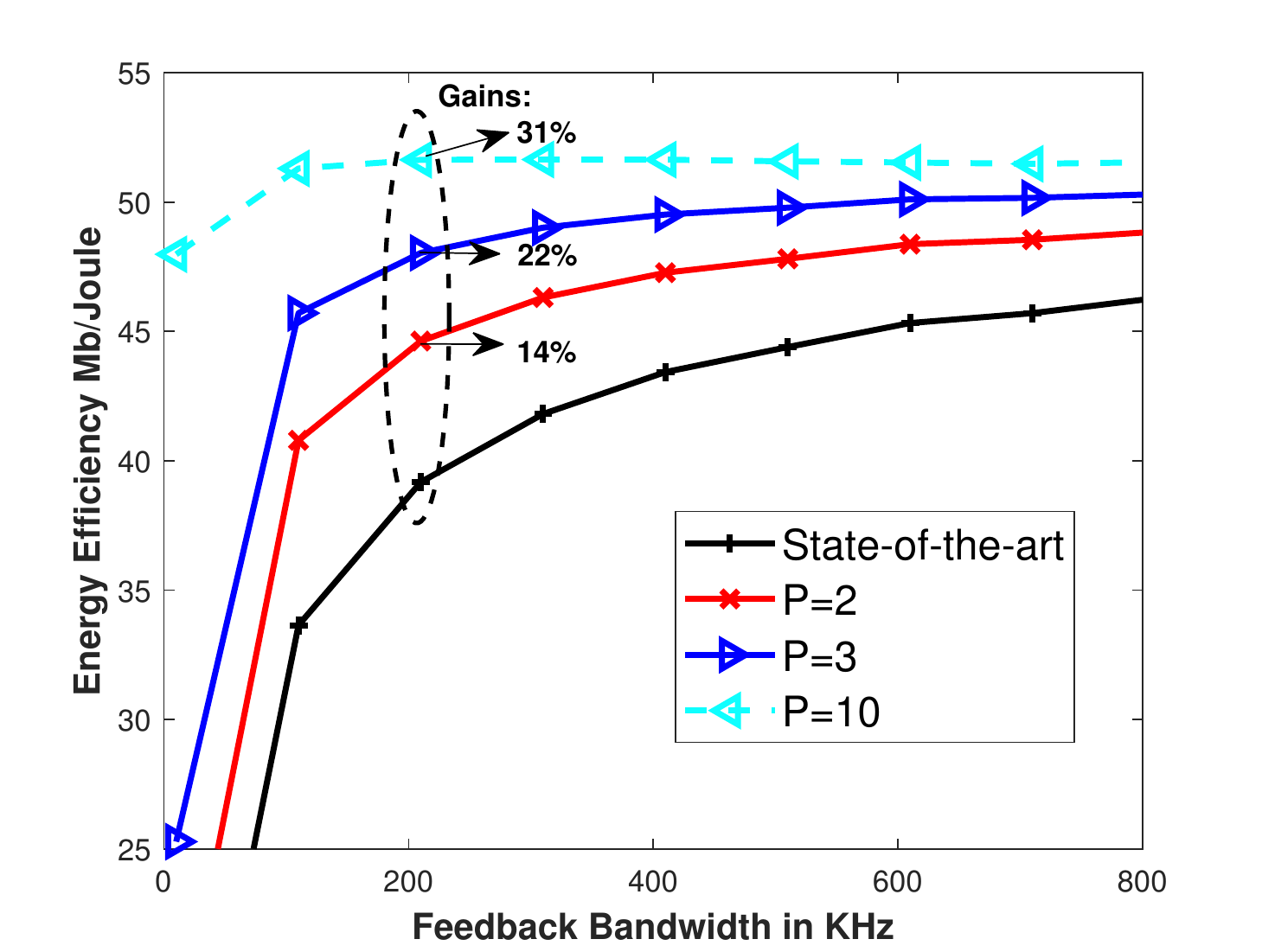}
	\caption{\ac{EE} performance of the state-of-the-art and the proposed method.}
	\label{fig:EE_vary_bf_fix_M}
\end{subfigure}
\caption{\ac{SE} and \ac{EE} performance of the proposed method varying the feedback bandwidth, with $N=1024$, $M_R=M_T=16$, $b^{(p)}_{\text{F}} = b_{\text{F}} = 3$ bits, for $p = 2,3,10$, for a Rician factor $K=10$ dB. }
\label{fig:SE_EE_fix_M}
\end{figure*}
\begin{figure*}[!t]
\centering
\begin{subfigure}{.48\linewidth}
  \centering
\includegraphics[scale=0.6]{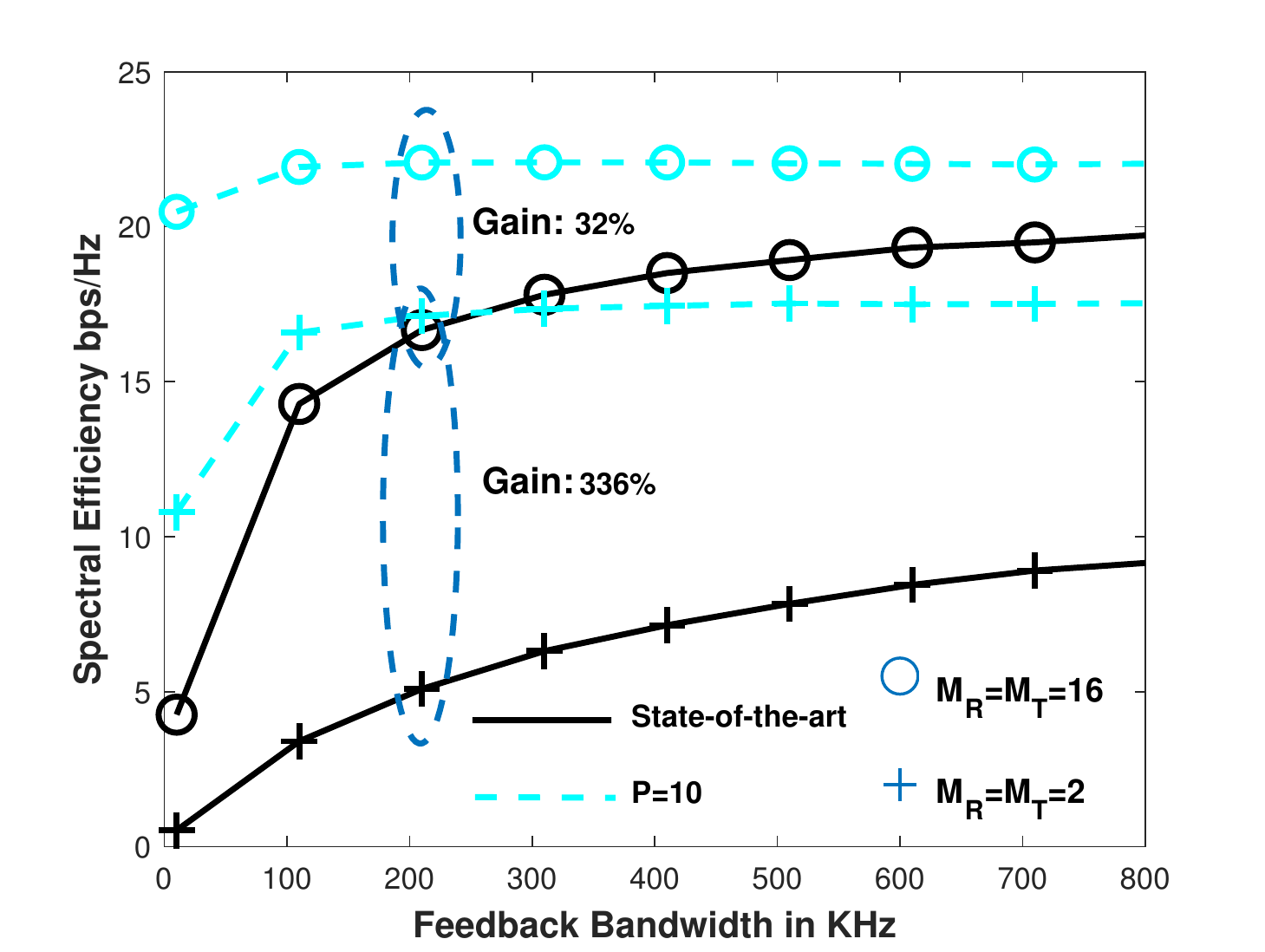}
\caption{\ac{SE} performance of the state-of-the-art and the proposed method..}
	\label{fig:SE_vary_bf_vary_M}
\end{subfigure}%
\hspace{0.2cm}
\begin{subfigure}{.48\linewidth}
  \centering
\includegraphics[scale=0.6]{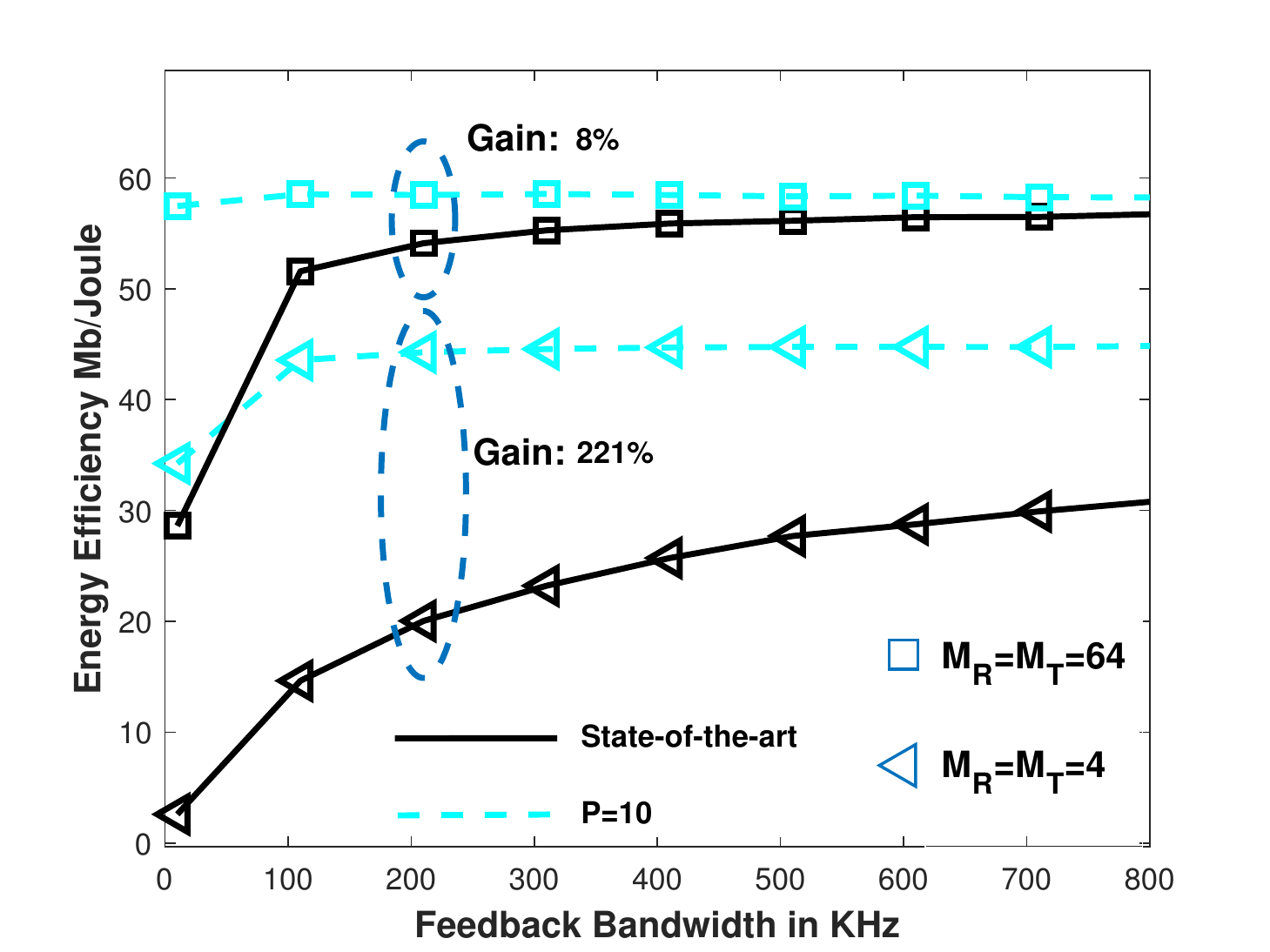}
	\caption{\ac{EE} performance of the state-of-the-art and the proposed method.}
	\label{fig:EE_vary_bf_vary_M}
\end{subfigure}
\caption{\ac{SE} and \ac{EE} performance of the proposed method varying the feedback bandwidth, with $N=1024$, $b^{(p)}_{\text{F}} = b_{\text{F}} = 4$ bits, for $p = 2,3,10$, for a Rician factor $K=10$ dB. }
\label{fig:SE_EE_vary_M}
\end{figure*}

\begin{figure*}[!t]
\centering
\begin{subfigure}{.48\linewidth}
  \centering
\includegraphics[scale=0.62]{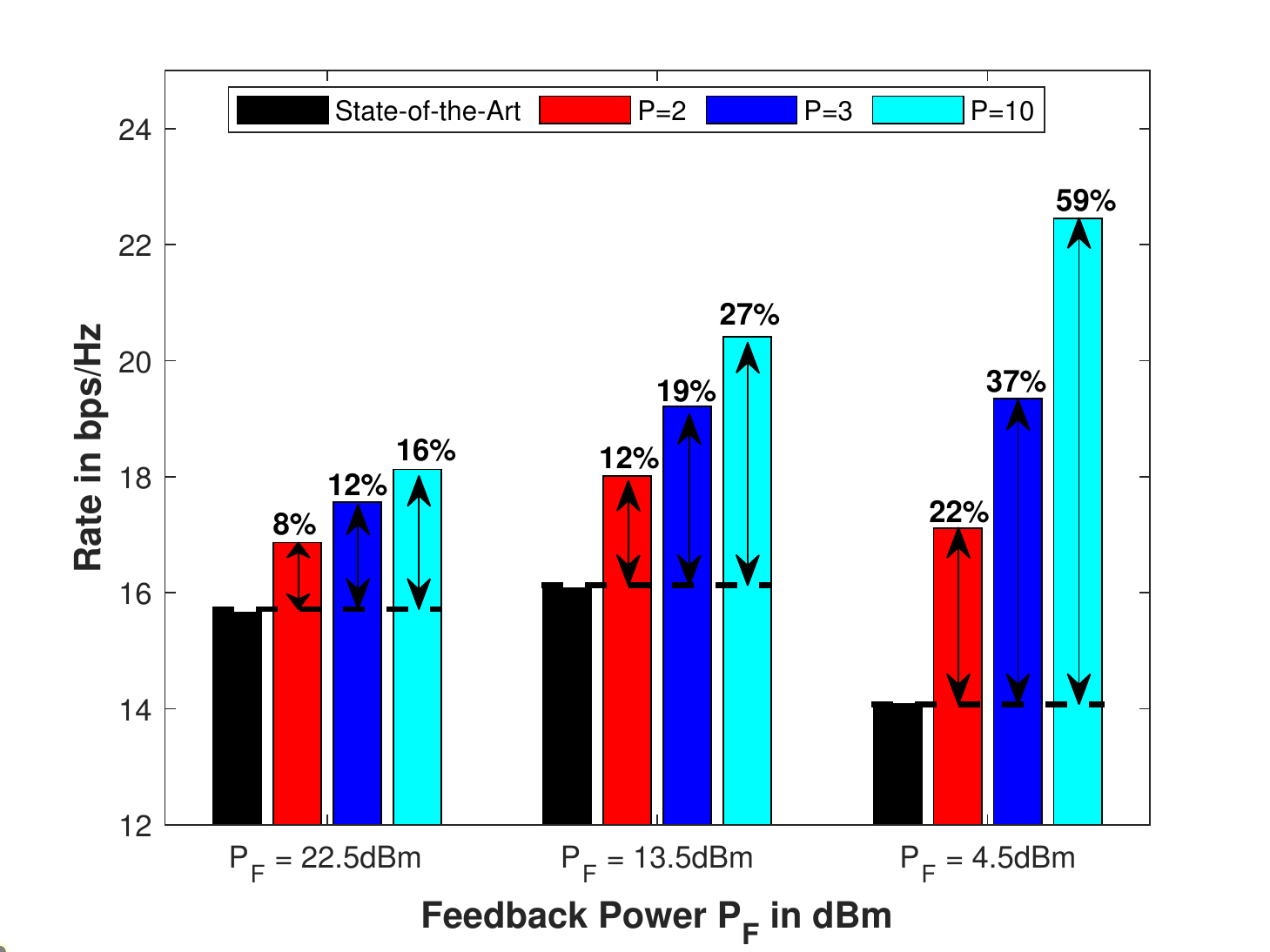}
\caption{\ac{SE} performance of the state-of-the-art and the proposed method.}
		\label{fig:SE_vary_pf}
\end{subfigure}%
\hspace{0.2cm}
\begin{subfigure}{.48\linewidth}
  \centering
\includegraphics[scale=0.62]{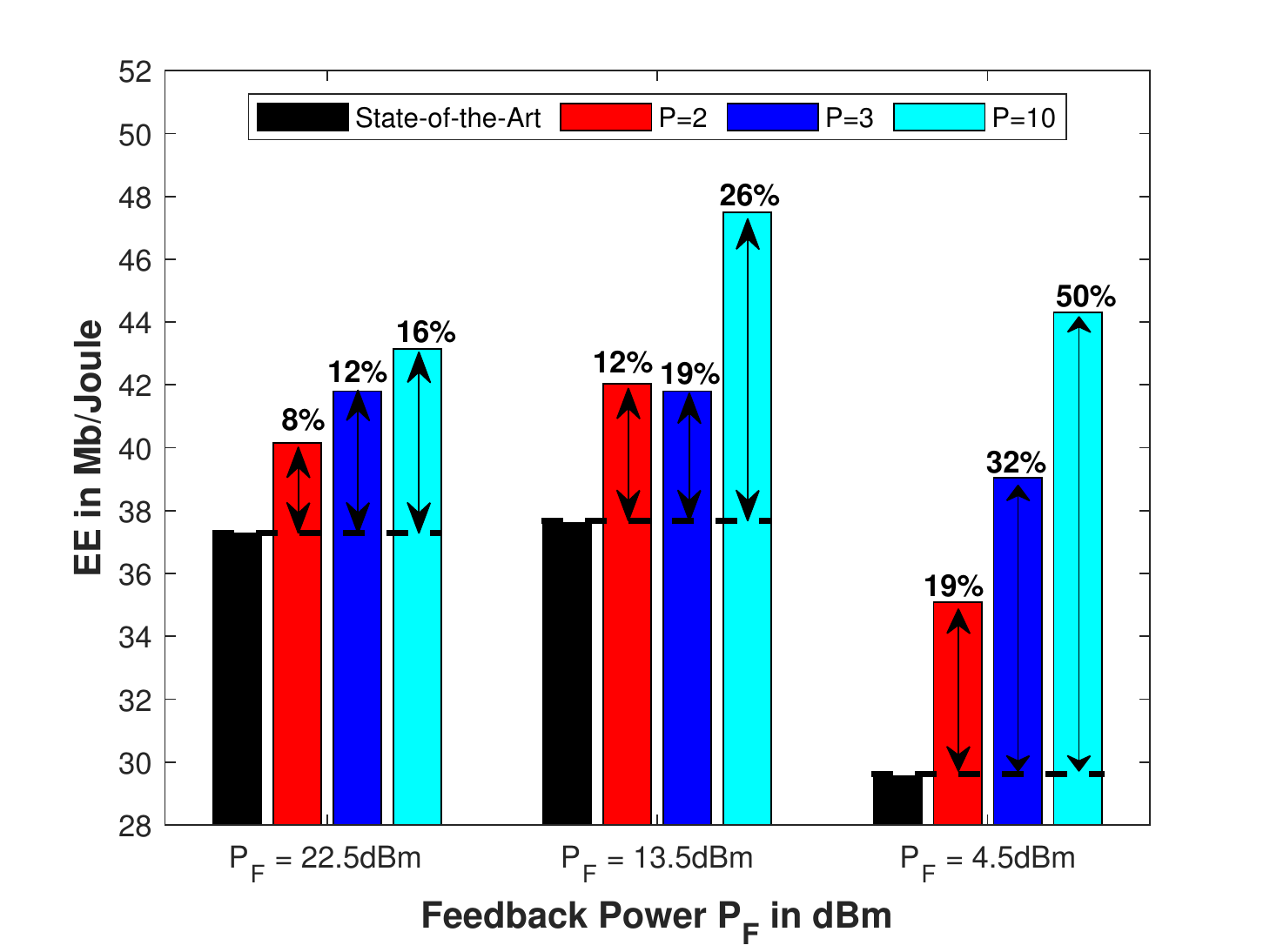}
	\caption{\ac{EE} performance of the state-of-the-art and the proposed method.}
	\label{fig:EE_vary_pf}
\end{subfigure}
\caption{\ac{EE} and \ac{SE} performance of the proposed method varying the feedback power, with $N=1024$, $M_R=M_T=2$, $b^{(p)}_{\text{F}} = b_{\text{F}} = 3$ bits, for $p = 2,3,10$, Rician factor $K=10$ dB. }
\label{fig:SE_EE_vary_pf}
\end{figure*}

\vspace{-2ex}
\subsection{Total System SE and EE Evaluation}
In this section, we evaluate the performance, in terms of \ac{SE} and \ac{EE}, of the proposed method by considering the total system rate, i.e., taking into account the channel estimation procedure duration and the IRS phase-shift feedback duration. To this end, we make use of the expressions given in (\ref{eq:rate_zap}) and (\ref{eq:power_zap}). The channel estimation period, in (\ref{eq:rate_zap}), is given as $\text{T}_{E} = (M_TN +1)T_0$, where $T_0= 0.8 \mu$ seconds denotes the duration of the pilot tones \cite{Zappone_Overhead_Aware}. The frame duration is given by $T = \text{T}_{PD}+ \text{T}_{\text{F}}$, where $\text{T}_{PD} = \text{T}_E + \text{T}_D$, is divided into $30\%$ for pilot transmissions ($\text{T}_E$) and $ 70\%$ for data transmission $\text{T}_{D}$. Regarding the power parameters of (\ref{eq:power_zap}), \textcolor{black}{we have} $P_{\text{E}} = P_0 (1 + NM_T)T_0$, where $P_0 = 0.8$ mW is the pilot tone power. Other parameter definitions can be found in Table \ref{tab:my-table}. The feedback channel  $g_{\text{F}}$ is generated from a circular symmetric complex Gaussian  distribution, normalized by $\sqrt{\beta_F} = \sqrt{\alpha_H} = \sqrt{\alpha_G}$ to account for the effects of pathloss and shadowing, as given in Table \ref{tab:my-table}.  In our next experiments, we assume $K=10$ dB, $N=1024$. For the proposed method, we consider the PARAFAC-IRS model with $R=1$. As for the number of factors, we study three configurations, with $P=2, (\ma{N}_2 = [512, 2]$), $P=3$ ($\ma{N}_{3} = [256, 2, 2]$) and $P=10$ ($\ma{N}_{10} = [2, \ldots, 2] \in \bb{R}^{10 \times 1}$.

\begin{table}[!t]
\centering
\resizebox{0.48\textwidth}{!}{%
\begin{tabular}{|l|l|l|l|l|}
\hline
$P_{\text{max}}$/$P_{c,0}$/$P_{c,n}$ & $B_{\text{max}}$ & $N_0$ & $\alpha_H$/$\alpha_G$ & $\mu$/$\mu_{\text{F}}$ \\ \hline
    $45$/$45$/$10$ dBm     &   $100$ MHz          &   $-174$ dBm/Hz    &        $110$/$110$ dB                   &    $1$/$1$  \\  \hline     
\end{tabular}%
}
\caption{}
\label{tab:my-table}
\end{table}

Figs.\ref{fig:SE_EE_fix_M} and \ref{fig:SE_EE_vary_M}, we analyze the total \ac{SE} and \ac{EE} of the proposed method  with the state-of-the-art \cite{Zappone_Overhead_Aware}, by varying the feedback bandwidth $B_{\text{F}} = B_{\text{max}} - B$, \textcolor{black}{where $B_\text{max}$ is the total available bandwidth given in Table \ref{tab:my-table}. As shown,} in Figs. \ref{fig:SE_vary_bf_fix_M} and \ref{fig:EE_vary_bf_fix_M}, when  the number of factors increases the feedback duration reduction pays off in the total system \ac{SE} and \ac{EE}.  The proposed method achieves a gain in the \ac{SE} of $32\%$ for $P=10$, $20\%$ for $P=3$, and $14\%$ for $P=2$,  over the state-of-the-art, considering the $B_{\text{F}} = 200$ kHz, with a similar gain in the \ac{EE}. 

In Figs. \ref{fig:SE_vary_bf_vary_M} and \ref{fig:EE_vary_bf_vary_M} we compare the proposed PARAFAC-IRS model under $R=1$ and $P=10$ with the state-of-the-art by varying the number of antennas. In this case, we observe that, for a feedback bandwidth $B_{\text{F}} \leq 200$ kHz, the proposed factorization with a $2 \times 2$  setup outperforms (in terms of \ac{SE} and \ac{EE}) the state-of-the-art one under the $4 \times 4$ setup, while presenting the same performance than the state-of-the-art one under the $16 \times 16$ setup. Finally, Figs. \ref{fig:SE_vary_pf} and \ref{fig:EE_vary_pf} show the \ac{SE} and \ac{EE} performances of the proposed method as a function of the feedback power $p_{\text{F}}$, with $p_{\text{F}} = P_{\text{max}} - p_{\text{TX}}$. We notice that the proposed configurations provide the best results in all scenarios.

{\color{black}
To summarize the results illustrated in Figs.\ref{fig:SE_EE_fix_M}-\ref{fig:SE_EE_vary_pf}, we conclude that the proposed tensor-based \ac{LRA} IRS phase-shift factorization models allows to reduce the number of phase-shifts to be conveyed to the IRS-controller, which significantly reduces the feedback overhead, resulting in \ac{SE} and \ac{EE} performance enhancements. In addition, our approach reaches similar performance to the non-factorized IRS, especially in moderate/strong \ac{LOS} scenarios, as it can be seen in Figs. \ref{fig:ALS_vs_HOSVD}-\ref{fig:fix_control_P_2}.
}
\textcolor{black}{From a system-level viewpoint, the network can resort to the proposed overhead-aware IRS model to increase the feedback periodicity, i.e., by providing more frequent feedback, which is crucial in fast time-varying channels, where the IRS should be reconfigured more frequently to follow the environment changes. Moreover, the proposed IRS factorization methods allow the network to multiplex more IRS phase-shifts in the same feedback channel, which is useful to accommodate multi-user IRS-assisted communications. 
}

\section{Conclusions and perspectives}\label{Sec:conclusions} 

In this paper, we   proposed two IRS phase-shift feedback overhead-aware methods based on tensor signal processing, namely, PARAFAC-IRS and Tucker-IRS. We \textcolor{black}{showed that the proposed methods significantly reduce  the IRS phase-shift feedback overhead, compared to the state-of-the-art approach, where the IRS phase shifts are not factorized. The PARAFAC-IRS method is preferable in the case of moderate/strong \ac{LOS} scenarios,  achieving a spectral efficiency that is close to that of the state-of-the-art, while providing a feedback overhead reduction. \textcolor{black}{Moreover,} in \ac{NLOS} scenarios, the Tucker-IRS model achieves a higher data rate than the PARAFAC-IRS model at the expense of a higher feedback overhead. By controlling the factorization parameters, we showed how to trade off data rate for feedback-overhead, allowing the network controller to adapt the IRS factorization parameters to meet a determined quality of service. } 


{\color{black}
\appendices
\section{Channel Model}\label{App:channel_model}
We provide details on the channel models for $\ma{H}$ and $\ma{G}$, given in (\ref{eq:channel_H}) and (\ref{eq:channel_G}), respectively. As mentioned, the \ac{NLOS} components of $\ma{H}$ and $\ma{G}$ are modeled as  random channels with $\bb{E}[\ma{H_{\text{NLOS}}}^{\text{H}}\ma{H}_{\text{NLOS}}] = \ma{I}_{M_T}$ and $\bb{E}[\ma{G}_{\text{NLOS}}\ma{G}^{\text{H}}_{\text{NLOS}}] = \ma{I}_{M_R}$. Nonetheless, the \ac{LOS} components are given as
\begin{align*}
\ma{H}_{\text{LOS}} &= \alpha_H \ma{b}_{\text{IRS}} \cdot \ma{a}^{\text{H}}_\text{TX} \in \bb{C}^{N \times M_T}, \\
\ma{G}_{\text{LOS}} &= \alpha_G \ma{b}_{\text{RX}} \cdot \ma{a}^{\text{H}}_\text{IRS} \in \bb{C}^{M_R \times N}, 
\end{align*}
where $\alpha_H$ and $\alpha_G$ are the path-loss components of the TX-IRS and IRS-RX links, respectively. Assuming that the TX and the RX are equipped with \acp{ULA} with half-wavelength inter-element spacing, their steering vectors can be written as
\begin{align}
\ma{a}_\text{TX} = \left[1, e^{j\pi \text{sin}\theta_{\text{TX}}}, \ldots, e^{j\pi (M_T -1) \text{sin}\theta_{\text{TX}}} \right]^{\text{T}} \in \bb{C}^{M_T \times 1},  \\
\ma{b}_\text{RX} = \left[1, e^{j\pi \text{sin}\theta_{\text{RX}}}, \ldots, e^{j\pi (M_R -1) \text{sin}\theta_{\text{RX}}} \right]^{\text{T}} \in \bb{C}^{M_R \times 1},
\end{align}
where $\theta_{\text{TX}}$ and $\theta_{\text{RX}}$ are the \ac{TX} and \ac{RX} \ac{AOD} and \ac{AOA}, respectively, which are generated from a uniform random distribution with  $\{\theta_{\text{TX}},\theta_{\text{RX}}\} \in [-\pi, \pi]$. {\color{black} Since the IRS is a $2$-D panel, the steering vectors associated with arrival and departure angles can be factorized as the Kronecker product of horizontal and vertical component vectors, respectively, as follows:}
\vspace{-.5ex}
\begin{align}
\ma{b}_{\text{IRS}} &= \ma{b}^{(v)}_{\text{IRS}} \otimes  \ma{b}^{(h)}_{\text{IRS}} \in \bb{C}^{N_hN_v \times 1}, \\
\ma{a}_{\text{IRS}} &= \ma{a}^{(v)}_{\text{IRS}} \otimes  \ma{a}^{(h)}_{\text{IRS}} \in \bb{C}^{N_hN_v \times 1}, 
\end{align}
where $N = N_hN_v$, $\ma{b}^{(h)}_{\text{IRS}} \in \bb{C}^{N_h \times 1}$ and $\ma{b}^{(v)}_{\text{IRS}} \in \bb{C}^{N_v \times 1}$  are the \ac{AOA} steering vectors in the azimuth and elevation directions, respectively. Likewise, $\ma{a}^{(h)}_{\text{IRS}} \in \bb{C}^{N_h \times 1}$ and $\ma{a}^{(v)}_{\text{IRS}} \in \bb{C}^{N_v \times 1}$  are the \ac{AOD} steering vectors in the azimuth and elevation directions, respectively. 
\vspace{-1ex}
\begin{align*}
\ma{b}^{(h)}_{\text{IRS}} &= [1, e^{j\pi \text{sin}\psi^{\text{AOA}}_{\text{IRS}}\text{cos}\phi^{\text{AOA}}_{\text{IRS}}}, \ldots, e^{j\pi (N_h -1)\text{sin}\psi^{\text{AOA}}_{\text{IRS}}\text{cos}\phi^{\text{AOA}}_{\text{IRS}}}],  \\
\ma{b}^{(v)}_{\text{IRS}} &= [1, e^{j\pi \text{cos}\phi^{\text{AOA}}_{\text{IRS}}}, \ldots, e^{j\pi (N_h -1)\text{cos}\phi^{\text{AOA}}_{\text{IRS}}}],  \\
\ma{a}^{(h)}_{\text{IRS}} &= [1, e^{j\pi \text{sin}\psi^{\text{AOD}}_{\text{IRS}}\text{cos}\phi^{\text{AOD}}_{\text{IRS}}}, \ldots, e^{j\pi (N_h -1)\text{sin}\psi^{\text{AOD}}_{\text{IRS}}\text{cos}\phi^{\text{AOD}}_{\text{IRS}}}],  \\
\ma{a}^{(v)}_{\text{IRS}} &= [1, e^{j\pi \text{cos}\phi^{\text{AOD}}_{\text{IRS}}}, \ldots, e^{j\pi (N_h -1)\text{cos}\phi^{\text{AOD}}_{\text{IRS}}}],
\end{align*}
where $\phi^{\text{AOA}}_{\text{IRS}}$ and $\phi^{\text{AOD}}_{\text{IRS}}$ are the elevation angles of arrival and departure, while $\psi^{\text{AOA}}_{\text{IRS}}$ and $\psi^{\text{AOD}}_{\text{IRS}}$ are the azimuth angles of arrival and departure. The azimuth angles $\psi^{\text{AOA}}_{\text{IRS}}$ and $\psi^{\text{AOD}}_{\text{IRS}}$ \textcolor{black}{are} generated from a uniform random distribution with $\{ \psi^{\text{AOA}}_{\text{IRS}}, \psi^{\text{AOD}}_{\text{IRS}}\} \in [-\pi,\pi]$, while the elevation angles $\phi^{\text{AOA}}_{\text{IRS}}$ and $\phi^{\text{AOD}}_{\text{IRS}}$ \textcolor{black}{are} generated from an uniform random distribution with $\{\phi^{\text{AOA}}_{\text{IRS}},\phi^{\text{AOD}}_{\text{IRS}}\} \in [0,\pi/2]$.
}  

\bibliographystyle{IEEEtran}
\bibliography{addons/ref.bib}
\end{document}

%% file: acronym-irs.tex
\newacronym{2G}{2G}{second generation}
\newacronym{3G}{3G}{third generation}
\newacronym{4G}{4G}{fourth generation}
\newacronym{5G}{5G}{fifth generation}
\newacronym{B5G}{B5G}{beyond fifth generation}
\newacronym{6G}{6G}{sixth generation}
\newacronym{3GPP}{3GPP}{3$\text{rd}$~Generation Partnership Project}
\newacronym{LTE}{LTE}{long term evolution}
\newacronym{NR}{NR}{new radio}
\newacronym{LS}{LS}{least squares}

\newacronym{IRS}{IRS}{intelligent reconfigurable surface}
\newacronym{RIS}{RIS}{reconfigurable intelligent surface}
\newacronym{LIS}{LIS}{large intelligent surface}
\newacronym{SDS}{SDS}{software-defined surface}

\newacronym{D2D}{D2D}{device-to-device}
\newacronym{BS}{BS}{base station}
\newacronym{UE}{UE}{user equipment}

\newacronym{SU}{SU}{single-user}
\newacronym{MU}{MU}{multi-user}
\newacronym{SISO}{SISO}{single-input single-output}
\newacronym{MISO}{MISO}{multiple-input single-output}
\newacronym{SIMO}{SIMO}{single-input multiple-output}
\newacronym{MIMO}{MIMO}{multiple-input multiple-output}

\newacronym{CSI}{CSI}{channel state information}
\newacronym{LOS}{LOS}{line of sight}
\newacronym{NLOS}{NLOS}{non-line of sight}

\newacronym{QoS}{QoS}{quality-of-service}
\newacronym{SE}{SE}{spectral efficiency}
\newacronym{EE}{EE}{energy efficiency}
\newacronym{SINR}{SINR}{signal to interference plus noise ratio}
\newacronym{SNR}{SNR}{signal to noise ratio}

\newacronym{ProSe}{ProSe}{proximity services}
\newacronym{NSPS}{NSPS}{national security and public safety}

\newacronym{RRM}{RRM}{radio resource management}
\newacronym{MS}{MS}{mode selection}
\newacronym{RA}{RA}{resource allocation}
\newacronym{PC}{PC}{power control}

\newacronym{BCD}{BCD}{block coordinate descent}

\newacronym{RF}{RF}{radio frequency}
\newacronym{AWGN}{AWGN}{additive white Gaussian noise}
\newacronym{MRC}{MRC}{maximum ratio combining}

\newacronym{AF}{AF}{amplify-and-forward}
\newacronym{DF}{DF}{decode-and-forward}

\newacronym{TX}{TX}{transmitter}
\newacronym{RX}{RX}{receiver}
\newacronym{ALS}{ALS}{alternating least squares}
\newacronym{SVD}{SVD}{singular value decomposition}
\newacronym{HOSVD}{HOSVD}{high order singular value decomposition}
\newacronym{THOSVD}{THOSVD}{truncated high order singular value decomposition}
\newacronym{PARAFAC}{PARAFAC}{PARAllel FACtors}
\newacronym{AOD}{AOD}{angle of departure}
\newacronym{AOA}{AOA}{angle of arrival}
\newacronym{URA}{URA}{uniform rectangular array} 
\newacronym{ADR}{ADR}{achievable data rate}
\newacronym{NMSE}{NMSE}{normalized mean square error}
\newacronym{SER}{SER}{symbol error rate}
\newacronym{LRA}{LRA}{low-rank approximation}

\newacronym{ULA}{ULA}{uniform linear array}
\newacronym{mmWave}{mmWave}{milimiter-wave}
\newacronym{CS}{CS}{compressed sensing}
\newacronym{OFDM}{OFDM}{orthogonal frequency division multiplexing}